\documentclass[aps,prc,twocolumn,nofootinbib,amsmath,showpacs,superscriptaddress,floatfix]{revtex4-1}

\usepackage{graphicx}
\usepackage{dcolumn}
\usepackage{bm}
\usepackage{mathrsfs}
\usepackage{color}

\begin{document}

\preprint{APS/123-QED}

\title{Nuclear level densities and gamma-ray strength functions in samarium isotopes}

\author{F.~Naqvi}
	\email{fnaqvi@nd.edu}
	\email{Current address: Department of Physics and Astrophysics, University of Delhi, India}
	\affiliation{Department of Physics, University of Notre Dame, IN 46556-5670, USA}
\author{A.~Simon}
	\email{anna.simon@nd.edu}
	\affiliation{Department of Physics, University of Notre Dame, IN 46556-5670, USA}
\author{M.~Guttormsen}
	\affiliation{Department of Physics, University of Oslo, N-0316 Oslo, Norway}
    
\author{R.~Schwengner}
	\affiliation{Helmholtz-Zentrum Dresden-Rossendorf, 01328 Dresden, Germany}
\author{S.~Frauendorf}
	\affiliation{Department of Physics, University of Notre Dame, IN 46556-5670, USA}
    
\author{C.S.~Reingold}
	\affiliation{Department of Physics, University of Notre Dame, IN 46556-5670, USA}	
    
\author{J.T.~Burke}
	\affiliation{Lawrence Livermore National Laboratory, Livermore, CA 94551, USA}	    
\author{N.~Cooper}
	\affiliation{Department of Physics, University of Notre Dame, IN 46556-5670, USA}
\author{R.O.~Hughes}
	\affiliation{Lawrence Livermore National Laboratory, Livermore, CA 94551, USA}	
\author{S.~Ota}
	\affiliation{Lawrence Livermore National Laboratory, Livermore, CA 94551, USA}  
\author{A.~Saastamoinen}
	\affiliation{Cyclotron Institute, Texas A\&M University, College Station, TX 77843, USA}

\date{\today}

\begin{abstract}
The $\gamma$-strength functions and level densities in the quasi-continuum of $^{147,149}$Sm isotopes have been extracted from particle-$\gamma$ coincidences using the Oslo method. The nuclei of interest were populated via (p,d) reactions on pure $^{148,150}$Sm targets and the reaction products were recorded by the Hyperion array. An upbend in the low-energy region of the $\gamma$SF has been observed. The systematic analysis of the $\gamma$SF for a range of Sm isotopes highlights the interplay between scissors mode and the upbend. Shell-model calculations show reasonable agreement with the experimental $\gamma$SFs and confirm the correspondence between the upbend and scissors mode.
\end{abstract}


\maketitle


\section{\label{sec:level1}Introduction}

The spectroscopic properties of excited nuclei provide information on the internal structure of these highly dense, many-body quantum systems.  Low-energy excitation regime is treated differently compared to the high-energy quasi-continuum region. In the latter, the quantities such as discrete energy levels are replaced by nuclear level densities (NLD) and transition probabilities are defined as $\gamma$-ray strength functions ($\gamma$SF) which are average reduced radiation or absorption probabilities at any given photon energy $E_{\gamma}$ \cite{bartholomew}.  Both of these observables also form important inputs for Hauser-Feshbach calculations predicting the astrophysical neutron capture rates \cite{feshbach}. Therefore,  a comprehensive understanding of NLD and $\gamma$SF is required for an insight on the astrophysical processes driving the synthesis of nuclei in our universe \cite{arnould,RevModPhys.83.157,goriely1}.  

The NLDs are often described by phenomenological analytical formulas built on the first principles of the Fermi gas model \cite{bethe}. However, due to the lack of experimental information on NLD, especially at high energies, the parametrization of the phenomenological models fails, giving rise to several microscopic approaches \cite{decowski, moretto, hillman,hilaire1,Hilaire2,alhassid,demetriou,hilaire3,goriely}. 

In order to explain the shape of the $\gamma$SF, phenomena such as giant electric dipole resonances are commonly adopted to fit the enhanced dipole transition probability at energies around 12-17 MeV \cite{dietrich}.  Below the neutron separation energy, an enhancement  in $\gamma$-ray strength is marked by excitation modes such as the $E1$ pygmy resonance ($E_\gamma$ $\sim$ 10 MeV) \cite{bracco,savran}, the $M1$ scissors mode in deformed nuclei ($E \sim$ 3 MeV) \cite{schiller}, or the $M1$ spin-flip resonance ($E \sim$ 8 MeV) \cite{heyde}.  The emergence of these contributions is well studied and explained based on sound theoretical calculations. However, a relatively recent observation of the strength enhancement in the energy range $E \leq 3-4$ MeV \cite{voivnov,guttormsen,wiedeking,larsen} does not  have  an affirmed origin yet.  Experiments involving the extraction of angular distributions established that this newly found low-energy {\it{upbend}} is of dipole nature \cite{larsen}. However, the information on its multipolarity is still elusive.  A recent polarization measurement of the photons originating in the $(p,p')$ reaction of $^{56}$Fe presented a preference for $M1$ character of the radiation in the region of enhancement \cite{jones}. This result is supported by the QRPA calculations \cite{Gor18} and the large-scale shell model calculations (LSSM) in $^{94-96}$Mo \cite{schwengner}, $^{56,57}$Fe \cite{brown} and $^{44}$Sc \cite{sieja} isotopes where the large B$(M1)$ strength at low energy, referred to as low-energy magnetic dipole radiation (LEMAR), is attributed to the reorientation of high-$j$ proton and neutron spins \cite{schwengner2}. This phenomenon is expected to appear near closed-shell nuclei having valence neutrons and protons in high-$j$ orbitals lying near to the Fermi surface. Recently, a more detailed theoretical investigation of the development of LEMAR across the $N= 28 - 50$ shell was performed for $^{60,64,68}$Fe nuclei \cite{schwengner2}. It was observed that the enhancement in the gamma-ray strength at $E <$ 3 MeV for the near closed-shell isotope $^{60}$Fe evolves into a bimodal structure comprising of a low-energy upbend and a scissors-like resonance at 3 MeV toward the mid-shell $^{64,68}$Fe nuclei. This theoretical result of the emergence of a bimodal structure in mid-shell nuclei was tested against the available experimental data on the $\gamma$SF of well-deformed $^{151,153}$Sm nuclei \cite{simon16}. Both the Sm isotopes exhibit well pronounced low-energy upbend and a bump at $\sim$ 3 MeV corresponding to the scissors mode. 

\begin{figure}[!ht]
\centering
\includegraphics[width=\columnwidth]{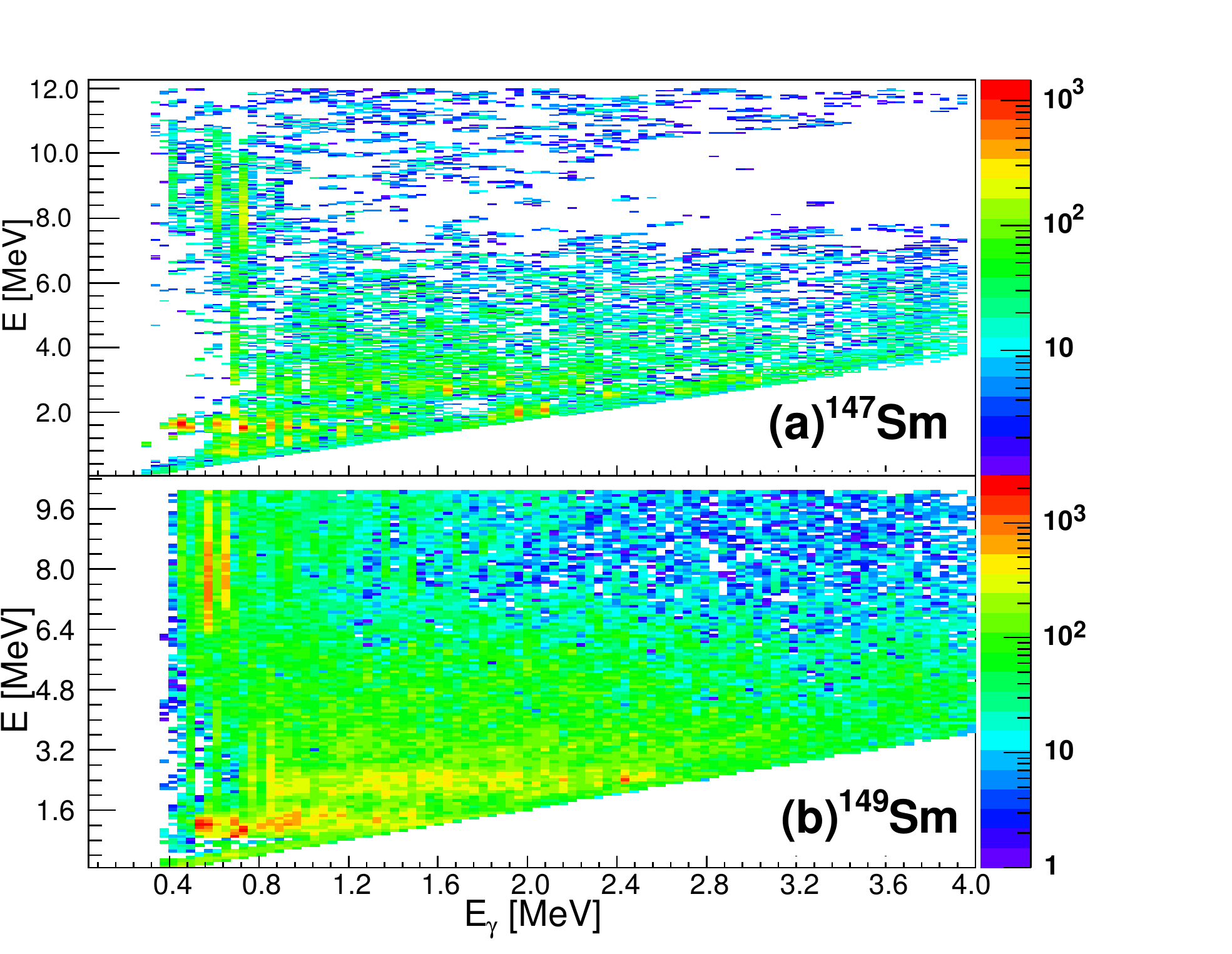}
\caption{\label{primary}First-generation (primary) $\gamma$-ray matrices for $^{147}$Sm (a) and $^{149}$Sm (b).}
\end{figure}

While the strength of the {\it{upbend}} and the scissors mode is a small contribution to the $\gamma$SF, it has a significant impact on capture and photodissociation reaction rates. TALYS calculations shown in Ref. \cite{simon16} highlight the profound effect of the observed low-energy strength enhancement on the neutron capture rates. An increase of 3 orders of magnitude in the rates is predicted for Sm isotopes lying at the neutron drip line provided a similar enhancement exists in that region. Measuring the $\gamma$SF in nuclei close to the neutron drip line is still a far-fetched goal, however a systematic study of the evolution of low-energy {\it{upbend}} in stable members of an isotopic chain is required to have a clear picture of the conclusions made in Ref. \cite{schwengner2} and to further extrapolate the properties of the $\gamma$SF to the less explored neutron-rich regions.  In this paper, the systematic study of the evolution of the $\gamma$SF at low energies was extended to $^{147,149}$Sm nuclei which are closer to the $N = 82$ shell. 

\section{Experimental procedure}

The experiment was performed at the Cyclotron Institute of Texas A\&M University, where two 98(1)\% isotopically enriched samarium targets, $^{148}$Sm and $^{150}$Sm,  0.8 mg/cm$^2$ and 1.1 mg/cm$^2$ thick, respectively, were bombarded by a 1.0~nA of 28~MeV proton beam from the K-150 cyclotron. The reaction products were detected by the Hyperion array  \cite{hughes17} that consists of 12 HPGe Clover-type $\gamma$-ray detectors combined with $\Delta E - E$ STARS telescope for charged particle identification and energy measurement. 

The telescope comprised two segmented silicon detectors, 140~$\mu$m ($\Delta E$) and 1000~$\mu$m ($E$) thick. Each of the detectors was a disk, 72~mm in diameter, with an 22~mm in diameter opening for the beam in the center. The disk was divided into 24 concentric 1~mm wide rings and into 8 segments in the angular direction. The $\Delta E-E$ system was placed 18~mm behind the target, providing an angular coverage for particle detection of 30-58 degrees. The design of the telescope allowed for identification of the light ion charged particle reaction products (protons, deuterons and tritons) and an energy resolution of 130~keV FWHM for detected deuterons.

The clover $\gamma$-ray detectors were positioned approximately 21~cm from the target at 45, 90, and 135 degrees with respect to the incident beam axis. Using standard $\gamma$-ray calibration sources, an energy resolution of 2.6 keV and 3.5 keV FWHM was obtained at 122 keV and 963 keV, respectively. The absolute photopeak efficiency of the Clover array was measured to be $\sim$10\% at 130 keV \cite{hughes17}. Only the $\gamma$ rays coincident with a particle were recorded, which provided data required to build the particle-$\gamma$ matrices for the Oslo method. The current study focused on two reactions: $^{148,150}$Sm(p,d$\gamma$)$^{147,149}$Sm.
\begin{figure}[!h]
\includegraphics[width=\columnwidth, clip, trim=0cm 0cm 1.8cm 0cm]{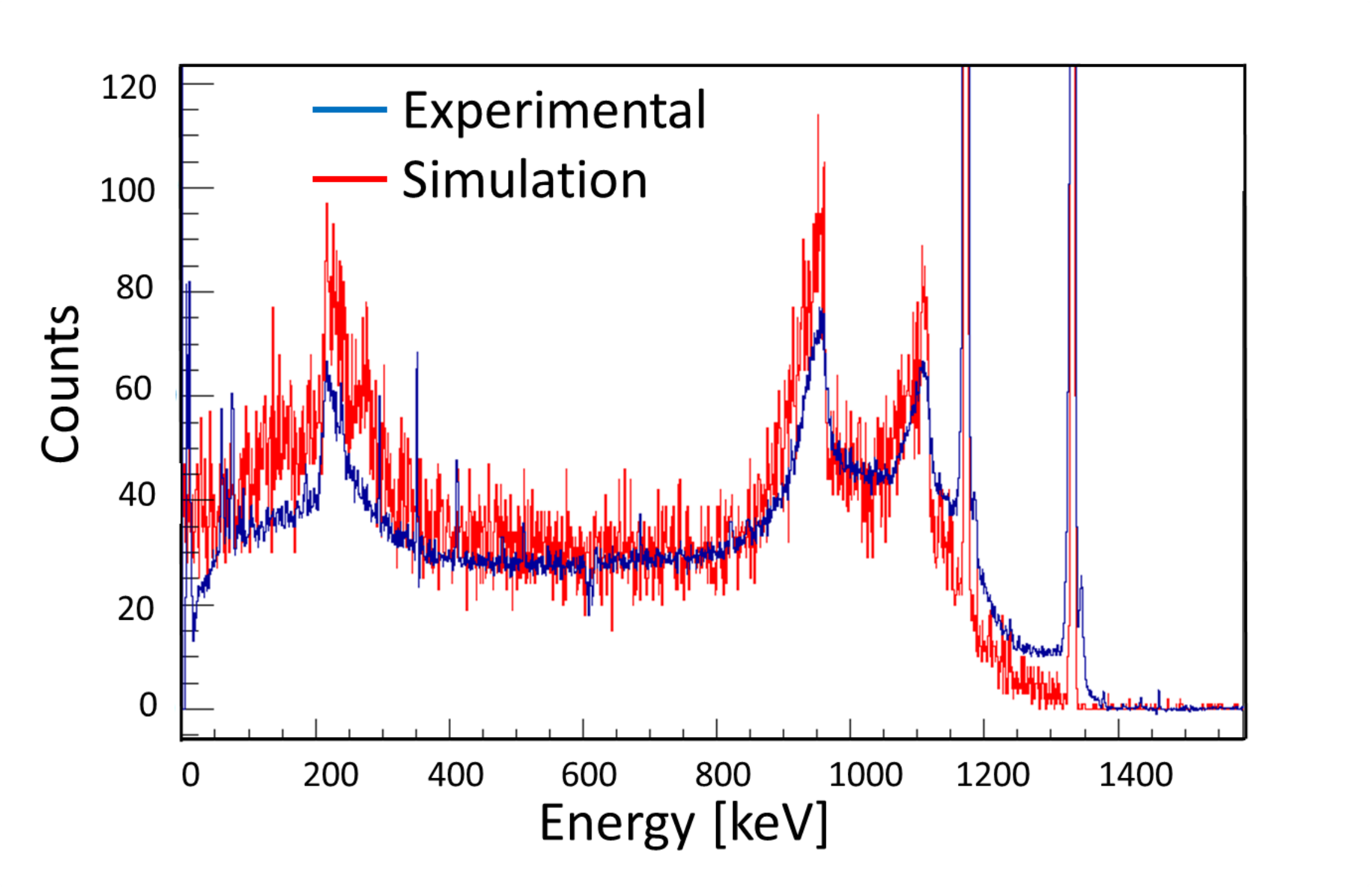}
\caption{\label{comp_simulation}A comparison of experimental  $\gamma$-ray spectrum of a clover detector with GEANT4 simulation for $^{60}$Co source.}
\end{figure}
\section{Analysis and Results}

\begin{figure*}[!ht]
\includegraphics[width=0.8\textwidth, clip, trim=0cm 0cm 1.8cm 0cm]{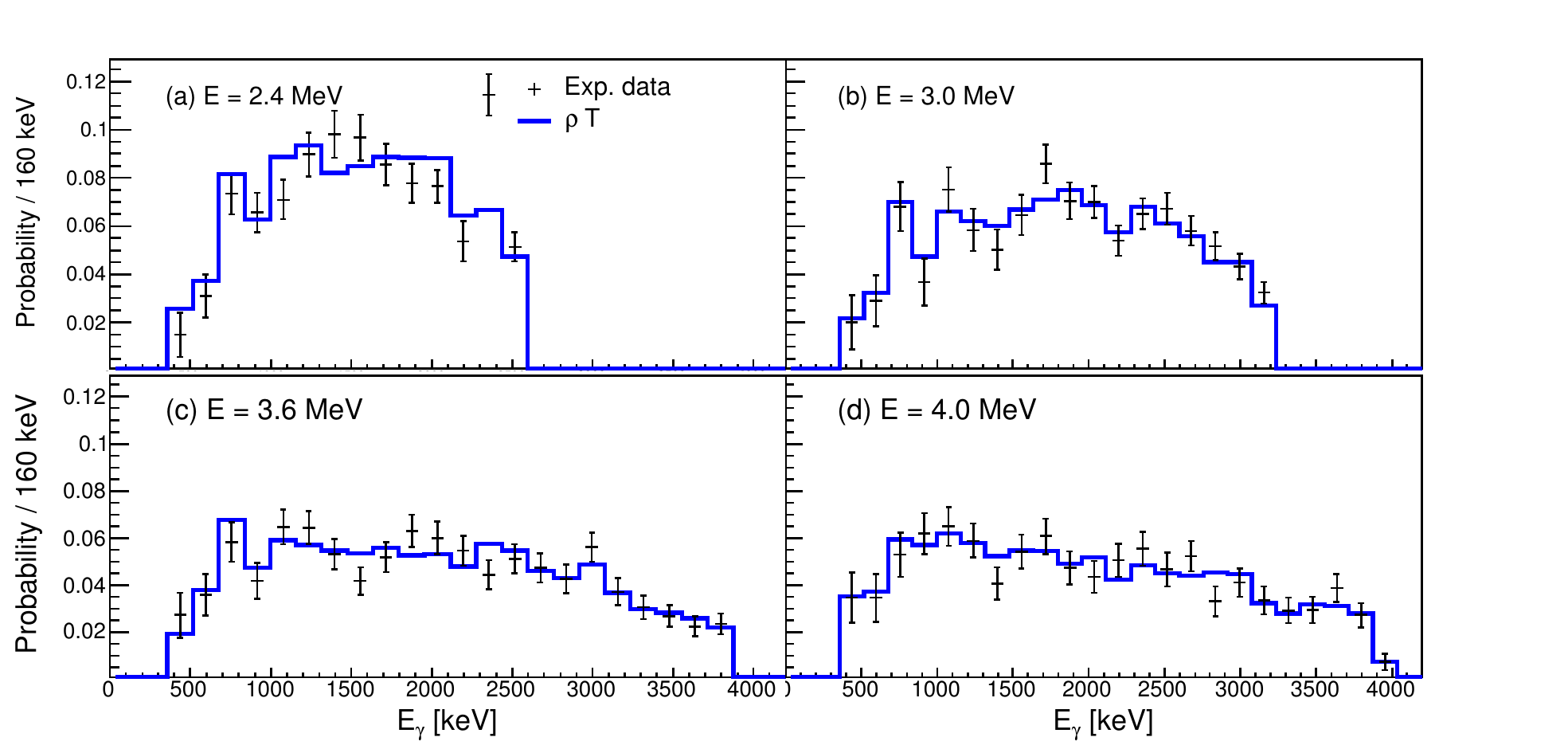}
\caption{\label{comparison}A comparison of experimental primary $\gamma$-ray spectra (crosses) for four excitation energies and the product of $\rho(E_f)$ and  $\mathscr{T}(E_{\gamma})$ obtained from the $\chi^{2}$ fit (solid lines) of $P(E,E_\gamma)$ in $^{149}$Sm. The fit is performed for the entire first-generation matrix and as shown here, works well for different subsets of excitation energies.}
\end{figure*}

To extract the NLD and $\gamma$SF from particle-$\gamma$ coincidence data, the Oslo method was used as the analysis technique \cite{oslo_method}.  This procedure relies on the fact that the $\gamma$ rays emitted in the first step of a decay cascade contain information about the level density and the $\gamma$-ray strength function. Therefore, the distribution of these first-generation, or the primary, $\gamma$ rays can be used to extract the functional form of the NLD and $\gamma$SF. The first step in obtaining a first-generation $\gamma$-ray distribution is to construct an excitation energy vs. $\gamma$-ray energy matrix which is then unfolded to correct for the efficiency of  the clover detector array.  For this purpose, response functions of the HPGe clovers were simulated for $\gamma$-ray energies up to 10 MeV with the Geant4 package \cite{geant4}. The unfolding procedure is an iterative process in which the shape of the Compton background, the single and double escape peak, and the annihilation peaks is estimated and subtracted from the observed spectrum to get the full energy $\gamma$-ray spectrum \cite{unfolding}. A comparison of experimental $\gamma$-ray spectrum of one clover detector with simulation for a $^{60}$Co source is presented in Fig. \ref{comp_simulation}.
The resulting unfolded $E$ vs. $E_\gamma$ matrix is divided into excitation-energy bins $i$ and $\gamma$ spectrum $f_{i}$ is projected for each of these bins. The spectra $f_{(j < i)}$ for the underlying bins $j$ consist of all the $\gamma$ rays in $f_{i}$ except the ones emitted first in the cascade. Thus, the primary $\gamma$-ray spectrum $h_{i}$ for each bin $i$ is obtained iteratively by subtracting $f_{i}$ and the weighted sum of all the spectra from underlying bins as,
\begin{equation}
h_{i}  = f_{i} - g_{i},
\end{equation}
Where $g_i$ is given by,
\begin{equation}
g_{i} = \sum_{j} n_{ij} w_{ij}f_{ij}. 
\end{equation}
The factors $n_{ij}$ correct for the difference in  population cross sections of excited states and $w_{ij}$ correspond to the probability of decay from states in bin $i$ to states in bin $j$. The latter constitute the weighing function $W$ which becomes equal to the primary $\gamma$-ray spectrum as the convergence is reached. It has been proven that the final primary $\gamma$-ray spectrum is independent of the first estimate of the weighing function $W$ \cite{first-generation2}. A detailed description of the unfolding procedure and creation of first-generation $\gamma$ rays is provided in Refs. \cite{unfolding,first-generation}. 

\begin{figure}[!ht]
\centering
\includegraphics[width=\columnwidth]{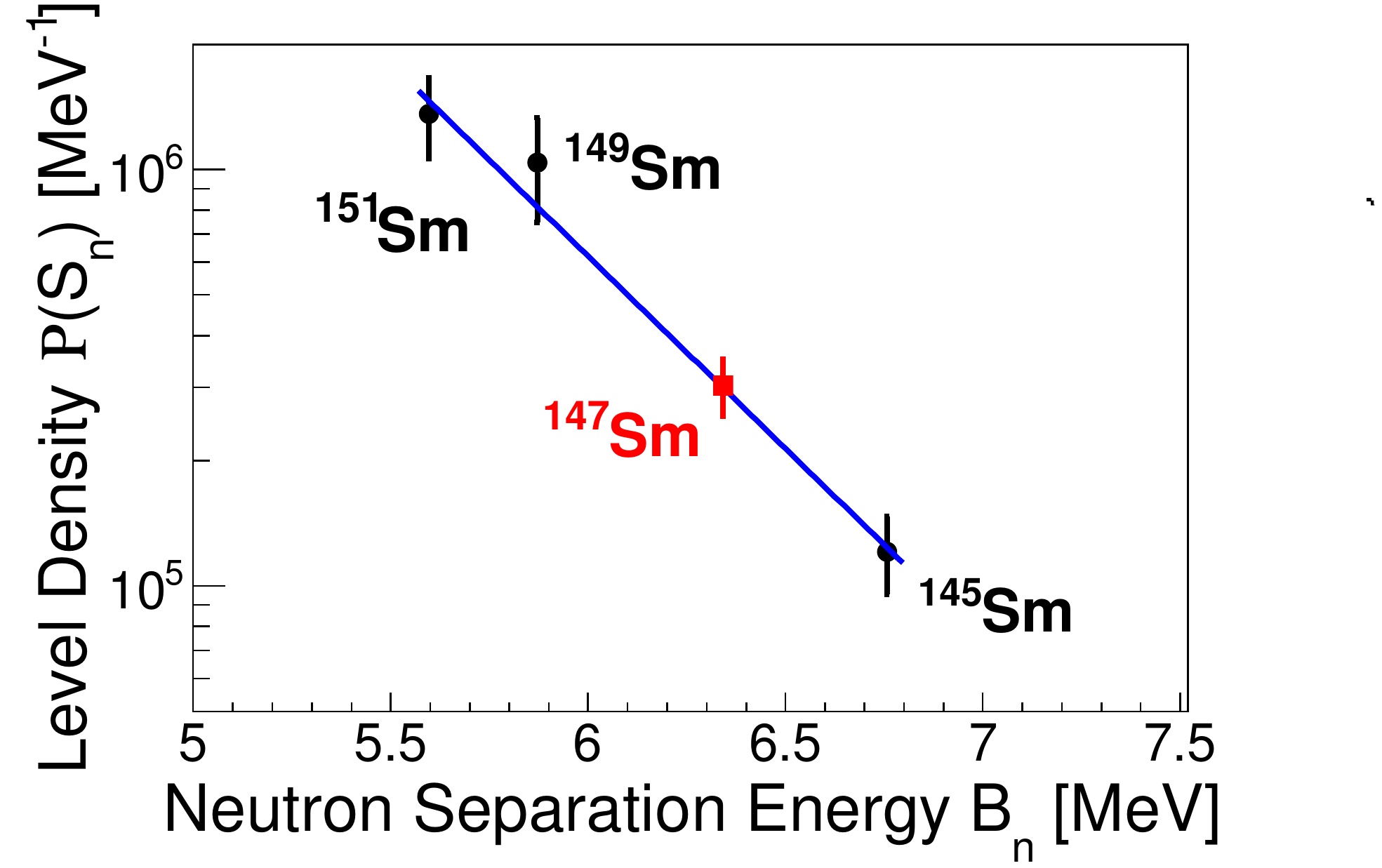}
\caption{\label{systematics}Systematics of level density as a function of neutron-separation energy $S_n$ for odd-$A$ Sm isotopes (black circles). The solid line corresponds to the linear fit of the $^{151,149,145}$Sm data points giving an estimate of the level density in $^{147}$Sm (red square). A reduction factor of 0.89 is used for all these $\rho(S_{n})$ calculated using the parameters given in  \cite{mughabghab}. }
\end{figure}

\begin{figure*}[!ht]
\centering
\includegraphics[width=1\textwidth]{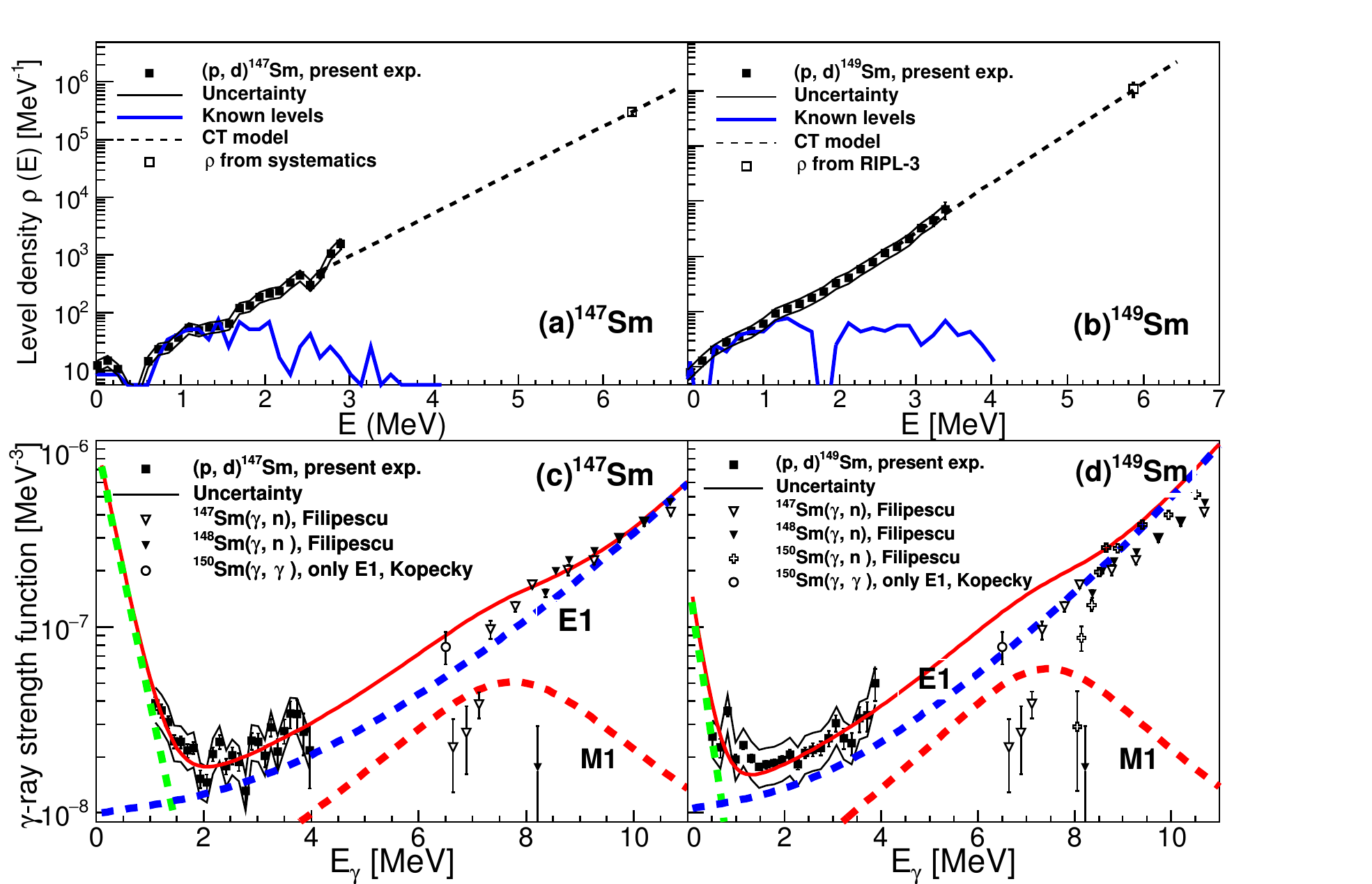}
\caption{\label{lvl_strength}Level-density functions for $^{147}$Sm (a) and $^{149}$Sm (b). Experimental data are shown as black squares. The dashed line corresponds to the constant-temperature approximation extrapolating to $\rho(S_{n})$ (open squares). The solid line is the known level density in the low-energy discrete region. Experimentally deduced $\gamma$-ray strength functions are shown in (c) and (d). For comparison, the analytical approximations for the E1 strength (blue dashed lines) and M1 strength (red dashed lines) as well as their sums (red solid lines) are shown.}
\end{figure*}
\vspace{250px}
\begin{table*}[!tb]
\caption{Parameters used for normalizing experimentally deduced level density and $\gamma$-ray strength function for $^{147,149}$Sm from the current work and for $^{151,153}$Sm taken from \cite{simon16}.}
\label{table1}
 \addtolength{\tabcolsep}{8pt}
\begin{tabular}{cccccccc}
\hline
\hline
Nucleus     & $S_n$     & $\sigma(S_n)$     & $D_0$                 & $\rho(S_n)$           & $<\Gamma_\gamma(S_n)>$    &$T_{CT}$   & Shift Parameter\\
            & (MeV)     &                   & (eV)                  & ($10^{6}$ MeV$^{-1}$) & (meV)                     & (MeV)     & (MeV)                        \\
\hline
$^{147}$Sm  & 6.342     & 6.266             & 252(40)\footnote{Estimated from systematics.}        & 0.31(5)$^{a}$         & 62(6)$^{a}$               &   0.58    &   -0.66     \\
$^{149}$Sm  & 5.871     & 6.121             & 65(13)                & 1.04(29)              & 66.9(14)                  &   0.48    &   -0.43   \\
$^{151}$Sm  &5.597      & 6.15              & 46(8)                 & 1.66(44)              & 60(5)                     &   0.51    &   -1.37             \\
$^{153}$Sm  &5.868      & 6.31              & 46(3)                 & 1.75(36)              & 60(5)                     &   0.53    &   -1.41                 \\
\hline
\hline

\end{tabular}
\end{table*}

For the present data, experimental first-generation $\gamma$-ray matrices $P(E,E_\gamma)$ for $^{147}$Sm and $^{148}$Sm are shown in Fig. \ref{primary}(a) and Fig. \ref{primary}(b), respectively. 
For statistical $\gamma$ decay, the Brink-Axel hypothesis \cite{brink,axel} allows to represent the primary $\gamma$-ray matrix $P(E,E_\gamma)$ as the product of level density $\rho(E_f)$ at the final excitation energy and the $\gamma$-ray transmission coefficient $\mathscr{T}(E_{\gamma})$,
\begin{equation}
P(E,E_\gamma) \propto \rho(E_{f})\mathscr{T}(E_\gamma). 
\end{equation}
As the above relation holds good only for the statistical regime, a lower limit on the excitation energy and the $\gamma$-ray energy is necessary while extracting the $\rho(E_f)$ and $\mathscr{T}(E_{\gamma})$ from $P(E,E_\gamma)$. In this analysis, conditions on $E^{min} = 2.5$ MeV, $E^{max} = 4.0$ MeV and $E_{\gamma}^{min} = 500$ keV were employed for both the Sm nuclei. Experimental statistics in the high-energy region determines the maximum value for the excitation energy in the analysis. A comparison of the experimental primary $\gamma$-ray spectra projected for different excitation energies and the product of $\rho(E_f)$ and  $\mathscr{T}(E_{\gamma})$ obtained from a $\chi^{2}$ fitting routine of 
$P(E,E_\gamma)$ in $^{149}$Sm is presented in Fig. \ref{comparison}. An overall good agreement is obtained between the data and the fit. 
The solution obtained after the fitting is the product of $\rho(E_f)$ and  $\mathscr{T}(E_{\gamma})$ which is unique but the individual  quantities are not. There are many functional forms of $\rho(E_f)$ and  $\mathscr{T}(E_{\gamma})$ which can give the same product. Therefore, to get the final $\tilde{\rho}$ and  $\tilde{\mathscr{T}}$ as,
\begin{equation}\label{ld}
\tilde{\rho} = Ae^{\alpha E_{f}}\rho(E_{f}),
\end{equation}
\begin{equation}
\tilde{\mathscr{T}} = Be^{\alpha E_\gamma}\mathscr{T}(E_{\gamma}),
\end{equation}
the A, B and $\alpha$ coefficients need to be determined with the help of the known experimental data. 

\begin{figure*}[!ht]
\centering
\includegraphics[width=0.8\textwidth]{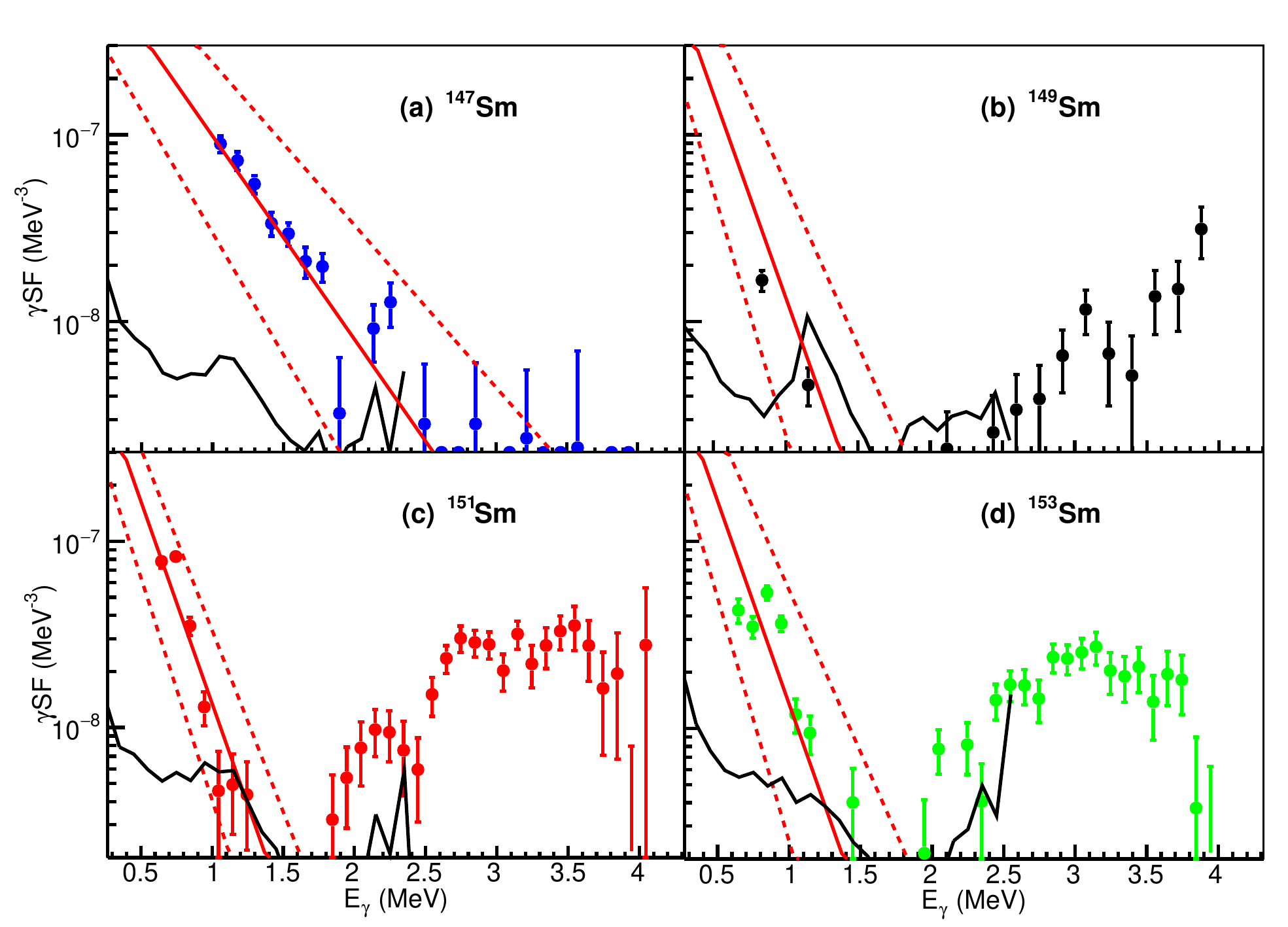}
\caption{\label{strengthall} $\gamma$-ray strength functions for all four Sm isotopes with the GDR contribution subtracted. Red solid lines indicate the fit to the upbend region, while the red dashed lines show the fit uncertainty. The results are compared with shell model calculations (black curve).}
\end{figure*}
\begin{table*}[!ht]
\caption{\label{table2}Parameters for resonances and the upbend for $^{147,149}$Sm isotopes from the current work and for $^{151,153}$Sm taken from \cite{simon16}.}
\addtolength{\tabcolsep}{-1pt}
\begin{tabular}{c|ccc|ccc|c|ccc|cc|ccc|c}
\hline\hline
Nucleus&\multicolumn{7}{|c}{Giant dipole 1 and 2 resonances }&\multicolumn{3}{|c}{Spin-flip M1}&\multicolumn{2}{|c}{Upbend}&\multicolumn{4}{|c}{Scissors resonance}\\
\cline{2-17}
    &$\omega_{E1,1}$&$\sigma_{E1,1}$&$\Gamma_{E1,1}$&$\omega_{E1,2}$&$\sigma_{E1,2}$&$\Gamma_{E1,2}$&$T_f$&$\omega_{\rm M1}$&$\sigma_{\rm M1}$&$\Gamma_{\rm M1}$  &$C$                  &$\eta$         &$\omega_{\rm SR}$  &  $\sigma_{\rm SR}$&  $\Gamma_{\rm SR}$  &   $B_{\rm SR}$ \\
           &(MeV)          &      (mb)     &      (MeV)    &   (MeV)       &     (mb)      &    (MeV)      &(MeV)&          (MeV)    &      (mb)        &         (MeV)  &(MeV$^{-3}$)&(MeV$^{-1}$)   & (MeV)    & (mb)     &   (MeV)    & ($\mu_N^2$) \\
\hline
$^{147}$Sm   &  13.8 &200 &3.8 &15.5 &230 &5.6 &0.55 &8.1 &2.3 &4.0 &$10(5)10^{-7}$  &3.2(10) &   -    &   -    &  -     &   -\\       
$^{149}$Sm   &  12.9 &180 &3.9 &15.7 &230 &6.5 &0.47 &7.7 &2.6 &4.0 &$20(10)10^{-7}$ &5.0(10) &    -   &   -    &   -    &   -\\
$^{151}$Sm   &  12.8 &160 &3.5 &15.9 &230 &5.5 &0.55 &7.7 &3.8 &4.0 &$20(10)10^{-7}$ &5.0(5)  &3.0(3) &0.6(2) &1.1(3) &7.8(34)  \\
$^{153}$Sm   &  12.1 &140 &2.9 &16.0  &232 &5.2 &0.45 &7.7 &3.3 &4.0 &$20(10)10^{-7}$ &5.0(10) &3.0(2) &0.6(1) &1.1(2) &7.8(20)  \\
\hline\hline
\end{tabular}
\end{table*}

To determine the parameters in Eq. \ref{ld}, the level density function $\rho(E_{f})$ is normalized to experimentally known discrete energy levels and the level density at neutron-separation energy $S_{n}$. As $^{147,149}$Sm are stable nuclei, the information on their level schemes for excitation energies up to $1 - 2$ MeV is comprehensive \cite{nndc}. The level density at $S_{n}$ is estimated from the spin-cutoff parameter $\sigma$ and the neutron-capture data which provides the  $s$-wave neutron-resonance spacing $D{_0}$. For $^{149}$Sm, $D_{0}$ value is taken from \cite{mughabghab} and $\sigma$ is provided by the NLD systematic study in Ref. \cite{egidy}. The latter is available for $^{147}$Sm, however, the $D_{0}$  value does not exist. Thus, a systematic study of level densities at $S_{n}$ for neighboring odd-$A$ Sm isotopes was performed. The calculated $\rho(S{_n})$ as a function of neutron separation energy are shown in Fig. \ref{systematics}. The level density for $^{147}$Sm is estimated by fitting an exponential function to the data points of odd-$A$ $^{151,149,145}$Sm. The higher-mass $^{153,155}$Sm isotopes are not included in the fit because of the variation observed in their trend. It is expected that $\rho(S_n)$ will increase as the atomic mass increases however, for $^{153,155}$Sm, a decreasing trend of $\rho({S_n})$ is observed which can be linked to the onset of deformation in these two isotopes. A similar behavior can be seen in deformed Dy isotopes \cite{Dy}. Table \ref{table1} lists the $D_0$ and spin-cutoff parameters used for normalizing the experimental level-density data in $^{147,149}$Sm nuclei. The parameters are consistent with the results obtained for the heavier Sm isotopes \cite{simon16}.

Once the level densities at low energies and at $S_n$ are determined, the slope of the experimental NLD curve is fixed by using the constant-temperature approximation (CT),
\begin{equation}\label{CT}
\rho_{CT}(E) = \frac{1}{T_{CT}} \exp \frac{E-E_{0}}{T_{CT}}.
\end{equation}
The CT fits shown in Fig. \ref{lvl_strength}(a) and Fig. \ref{lvl_strength}(b) yield constant-temperature and shift parameters as given in Table \ref{table1}. $T_{CT}$ of 0.58 and 0.48 MeV are obtained for $^{147,149}$Sm, respectively which are in accordance with the values reported in Ref. \cite{simon16} for heavier $^{151,153}$Sm isotopes.

The last step is to find the scaling parameters for the $\gamma$-ray transmission coefficient $\mathscr{T}(E_{\gamma})$. 
The average total radiative width $<\Gamma_\gamma>$ at $S_n$ needed for normalizing the $\mathscr{T}(E_{\gamma})$ was taken from \cite{mughabghab}. The normalization procedure is described in detail in \cite{schiller,voivnov} and the parameters are summarized in Table \ref{table1}.

The transmission coefficients can be converted to the dipole $\gamma$-ray strength function as:
\begin{equation}
f(E_\gamma) = \frac{1}{2\pi}\frac{\mathscr{T}(E_{\gamma})}{E^{3}_{\gamma}}.
\end{equation}
The resulting experimental $\gamma$SF are presented in solid squares in Fig. \ref{lvl_strength}(c) and \ref{lvl_strength}(d), respectively. Additionally, results from ($\gamma,n$) cross section measurements from Filipescu {\it{et al.}}  \cite{filipescu} are also shown for comparison. The dipole strength functions shown were calculated from the reaction cross section given in \cite{filipescu} and \cite{ripl3}:
\begin{equation}
f(E_{\gamma}) = \sigma(E_{\gamma}) / (3\pi^2 \hbar^2c^2 E_{\gamma}).
\end{equation} 
The combined data sets were then fitted with two generalized Lorentzians (GLOs) for the giant electric dipole resonance (GDR) as defined in RIPL-3 \cite{ripl3}. The $M1$ spin-flip resonance was fitted with a Lorentzian shape with estimates of parameters given in RIPL-3. 

The measured $\gamma$SFs shown in Fig. \ref{lvl_strength} show a distinct feature at low energies: an enhancement at energies below 2 MeV. This feature, an upbend, has previously been observed in deformed $^{151,153}$Sm isotopes \cite{simon16} in combination with the scissors mode at around 3 MeV. In the case presented here, the $^{147,149}$Sm isotopes are nearly spherical, thus the scissors mode is not present.

Following the procedure from \cite{simon16}, the upbend was fitted with:
\begin{equation}
\label{eqn:upbend}
f_{\rm upbend}(E_{\gamma}) = C\exp(-\eta E_{\gamma}).
\end{equation}
The results are shown as green dashed line if Fig. \ref{lvl_strength} (c) and (d) and the fit parameters are listed in Table \ref{table2}. The fit parameters obtained for $^{147,149}$Sm isotopes are consistent with those for $^{151,153}$Sm from \cite{simon16} as can be observed from Table \ref{table2}.

\begin{table}[!tb]
\caption{\label{table3} Total $B(M1)$ strength in the 0-5 MeV region calculated for the upbend and scissor components of the $\gamma$SF.}
\addtolength{\tabcolsep}{4pt}
\begin{tabular}{lcccc}
\hline\hline
Nucleus & $^{147}$Sm & $^{149}$Sm & $^{151}$Sm & $^{153}$Sm \\
$B(M1)_{tot} (\mu_N^2)$ & $9.8^{+16.7}_{-6.3}$ & $7.2^{+9.8}_{-4.2}$ & $8.0^{+9.8}_{-4.2}$ & $8.0^{+9.8}_{-4.2}$ \\
\hline\hline
\end{tabular}
\end{table}

In Fig. \ref{strengthall}, the $\gamma$-strength functions for all four Sm isotopes, $A$ = 147,149,151,153, are shown with the GDR component subtracted. Thus, only the upbend and the scissor components are present in the plots. The fit to the upbend, with uncertainties as listed in Table \ref{table2} is also shown. It can be seen that the strength of the scissors mode at about 3 MeV increases with the mass number. This is consistent with the deformation of the Sm isotopes in this region, which increases with the increasing number of neutrons. The $^{147}$Sm isotope is nearly spherical, thus the scissors mode is not present.

Within the uncertainty of the fit to the upbend region of the $\gamma$SF it is difficult to assess the trend of the $\gamma$SF as a function of $N$. However, for comparison with Ref.  \cite{schwengner2}, the total $B(M1)$ strength in the $E_\gamma$ region of 0 -- 5 MeV is obtained from a numerical integration of the strength function:
\begin{equation}
    B(M1)_{tot}=\frac{9}{16\pi} (\hbar c)^3 \sum f_{M1}(E_\gamma)\Delta E_\gamma.
\end{equation}
The resulting strengths are listed in Table \ref{table3} for all four Sm isotopes. The uncertainties listed in the table represent the maximum error in the $B(M1)_{tot}$ and were calculated by integrating the upper and lower limits of the fits to the upbend and scissors components based on the uncertainties in the fit parameters listed in Table \ref{table2}. The total strengths for all of the Sm isotopes are comparable and deviate by less than 13\% from the average value of $8.27^{+11.2}_{-4.7}$. This result is in a very good agreement with the predictions from \cite{schwengner2}.

\begin{figure*}[!ht]
\centering
\includegraphics[width=0.8\textwidth]{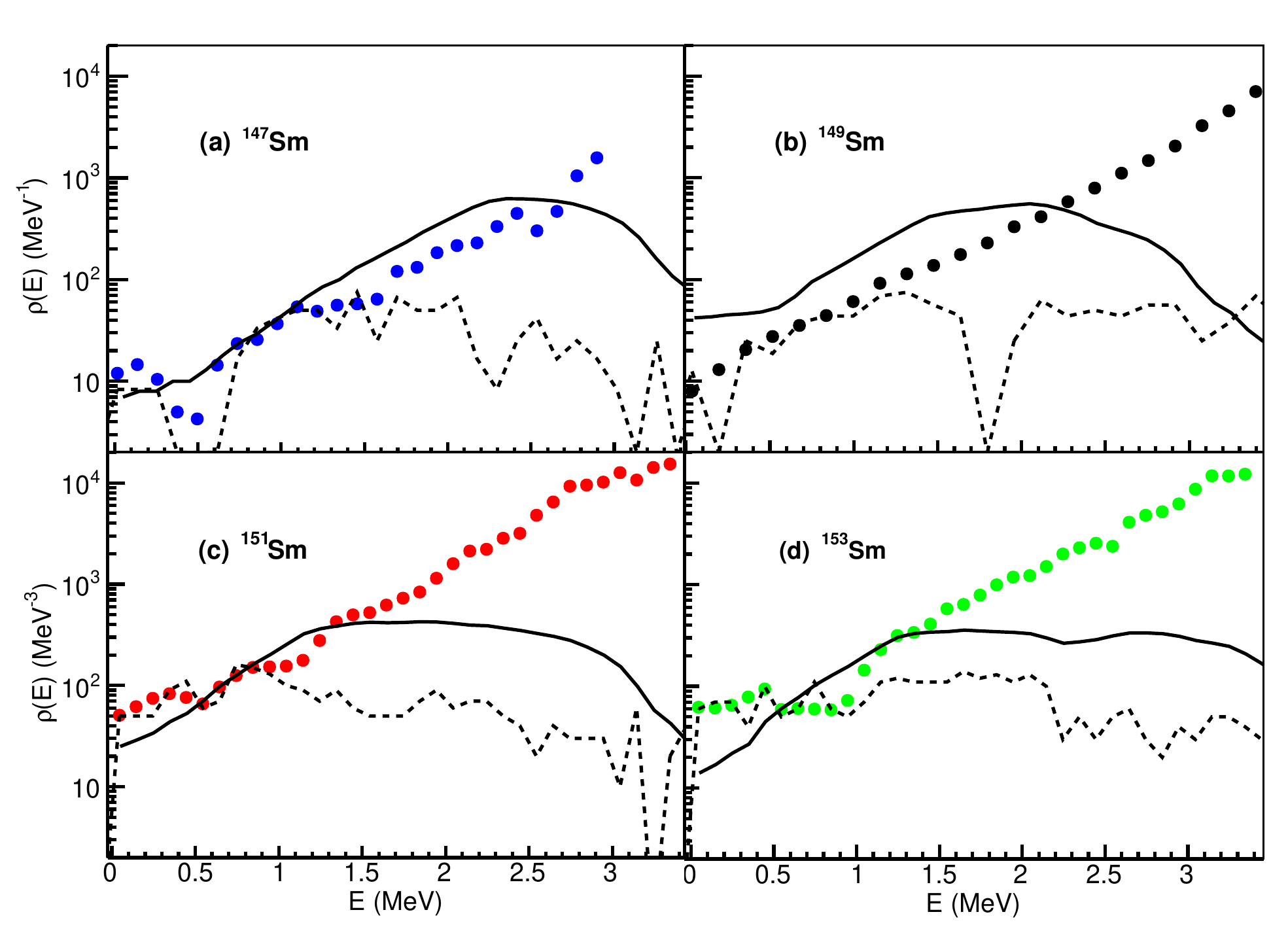}
\caption{\label{ldall} Level density functions for all four Sm isotopes: solid symbols - data extracted using the Oslo method, solid line - level density from shell-model calculations, dashed line - known levels.}
\end{figure*}

\section{Shell-model calculations}

The experimental results for the $\gamma$SFs of all four Sm isotopes were compared to predictions of shell-model calculations. The calculations were carried out in the jj56pn model space with the jj56pna Hamiltonian using the code NuShellX@MSU \cite{bro14Nu}. The model space included the
$(1g_{7/2}, 2d_{5/2}, 2d_{3/2}, 3s_{1/2}, 1h_{11/2})$ proton orbits and the $(1h_{9/2}, 2f_{7/2}, 2f_{5/2}, 3p_{3/2}, 3p_{1/2}, 1i_{13/2})$ neutron orbits relative to a $^{132}$Sn core. In the present calculations, two protons were allowed to be lifted to the $2d_{3/2}, 3s_{1/2}$, and $1h_{11/2}$ orbits, and two neutrons could be excited to the $2f_{7/2}, 3p_{3/2}$, and $2f_{5/2}$ orbits. The calculations of $M1$ strengths included the lowest 60 states each with spins of $J_i, J_f$ = 1/2 to 13/2. The range of spins populated in the reaction is based on the results from Cooper et al. \cite{cooper18}. Effective $g$ factors of $g^{\rm eff}_s = 0.7  g^{\rm free}_s$ were applied. The reduced transition strengths $B(M1)$ were calculated for all transitions from initial to final states with energies $E_i > E_f$ and spins $J_i = J_f, J_f \pm 1$. This resulted in more than 28,000 $M1$ transitions for each parity.

Strength functions were deduced according to
\begin{multline}
f_{M1}(E_\gamma,E_i,J_i,\pi) = \\
\frac{16\pi}{9} (\hbar c)^{-3} \overline{B}(M1,E_i \rightarrow E_f, J_i, \pi)\rho(E_i, J_i, \pi),
\end{multline}
where $E_\gamma = E_i - E_f$, $\overline{B}(M1)$ are averages in considered 
$(E_i,E_f)$ elements for given $J_i, \pi$, and $\rho(E_i, J_i, \pi)$ are level densities from the present calculations. The strength functions $f_{M1}(E_\gamma)$ were obtained by averaging step-by-step over $E_i$, $J_i$, and $\pi$.

The calculated level densities are included in Fig. \ref{ldall}. In contrast to the steadily increasing experimental NLD, the restricted number of levels causes a cut-off of the calculated NLD above about 3 MeV, which is the approximate energy of the highest of the 60 levels taken into account for each spin. As a consequence the theoretical curves saturate and bend over at 2.5, 1.5, 1.2, 1.2 MeV for $N$ = 85, 87, 89, 91, respectively, which is the signal of the missing levels. Only below these energies the calculated level densities can be compared with the experimental ones. There, the calculations follow the CT expression (\ref{CT}) with $T_{CT}\approx 0.5$ MeV in good agreement with the experimental values $T_{CT}$ in Table \ref{table2}. The average scale of the experimental level densities is well  reproduced 
 for $N$ = 85, 89, 91. It is overestimated by a  factor of two  for $N$ = 87. The calculations do not account for the details of the $E_x$ dependence of the level density, nor are they capable of reproducing the energies of the observed lowest levels. 
 

The calculated strength functions are included in Fig. \ref{strengthall}. Because of the high level density of the Sm isotopes, the highest of the included 60 levels at each spin appear around 3 MeV, and consequently the $\gamma$-ray energies reach up to about 2.5 MeV only. For $^{149,151,153}$Sm, the calculated strength functions reproduce the experimental ones in the range of 1 MeV $< E_\gamma <$ 2.5 MeV. They account for the development of the dip between the decreasing upbend and the starting scissors component with increasing deformation. The present calculations affirm the findings in Ref. \cite{schwengner2} that the shell-model calculations indicate the appearance of the scissors mode in deformed nuclei. The calculations substantially underestimate the strength below 1 MeV. In contrast to earlier calculations for lighter nuclei \cite{schwengner,schwengner2,brown,sieja2018,mid18} the shape deviates from the exponential form seen in the experiment. We attribute this discrepancy as well as the deviations of the calculated from experimental level densities to the restrictions enforced by the numerical effort. The missing strength for low energy $\gamma$ rays may signal that calculations do not account for transitions between closely spaced levels above the highest calculated ones.


\section{Summary}
Level densities and $\gamma$-ray strength functions were extracted from particle-$\gamma$ coincidence data for $^{147,149}$Sm nuclei using the Oslo method. As in the previous study of the Sm nuclei, the low-energy upbend in the $\gamma$SF has been observed at energies below 2 MeV. No structure that could be attributed to the scissors mode has been observed, which is consistent with the lack of deformation of the studied nuclei. The results of this work are consistent with the previous measurements of the statistical properties of the Sm nuclei \cite{simon16}. Moreover, the total M1 strength in the $\gamma$-ray energy range of 0-5 MeV remains fairly constant across the isotopic chain as it was predicted by Schwengner et al \cite{schwengner2}.
Shell model calculations for the lowest 60 levels for spin 1/2 - 13/2 were carried out, which reproduce the gross structure of the experimental level densities (exponential increase with excitation energy) and of the $\gamma$-ray strength functions (development of a minimum at a transition energy  of about 1.7 MeV caused by the emergence of a scissors resonance with the onset of deformation).

\section{Acknowledgments}
This work was supported by the U.S. Department of Energy No. DE-NA0002914, DE-NA0003780, DEFG02- 95ER-40934, DE-NA0003841 and by Lawrence Livermore National Laboratory under Contract No. DE-AC52-07NA27344. R.S. thanks B. A. Brown for his support in using the code NuShellX@MSU and acknowledges the cooperation of the Centers for High-Performance Computing of Technische Universit\"at Dresden and of Helmholtz-Zentrum Dresden-Rossendorf.


\begin{thebibliography}{49}%
\makeatletter
\providecommand \@ifxundefined [1]{%
 \@ifx{#1\undefined}
}%
\providecommand \@ifnum [1]{%
 \ifnum #1\expandafter \@firstoftwo
 \else \expandafter \@secondoftwo
 \fi
}%
\providecommand \@ifx [1]{%
 \ifx #1\expandafter \@firstoftwo
 \else \expandafter \@secondoftwo
 \fi
}%
\providecommand \natexlab [1]{#1}%
\providecommand \enquote  [1]{``#1''}%
\providecommand \bibnamefont  [1]{#1}%
\providecommand \bibfnamefont [1]{#1}%
\providecommand \citenamefont [1]{#1}%
\providecommand \href@noop [0]{\@secondoftwo}%
\providecommand \href [0]{\begingroup \@sanitize@url \@href}%
\providecommand \@href[1]{\@@startlink{#1}\@@href}%
\providecommand \@@href[1]{\endgroup#1\@@endlink}%
\providecommand \@sanitize@url [0]{\catcode `\\12\catcode `\$12\catcode
  `\&12\catcode `\#12\catcode `\^12\catcode `\_12\catcode `\%12\relax}%
\providecommand \@@startlink[1]{}%
\providecommand \@@endlink[0]{}%
\providecommand \url  [0]{\begingroup\@sanitize@url \@url }%
\providecommand \@url [1]{\endgroup\@href {#1}{\urlprefix }}%
\providecommand \urlprefix  [0]{URL }%
\providecommand \Eprint [0]{\href }%
\providecommand \doibase [0]{http://dx.doi.org/}%
\providecommand \selectlanguage [0]{\@gobble}%
\providecommand \bibinfo  [0]{\@secondoftwo}%
\providecommand \bibfield  [0]{\@secondoftwo}%
\providecommand \translation [1]{[#1]}%
\providecommand \BibitemOpen [0]{}%
\providecommand \bibitemStop [0]{}%
\providecommand \bibitemNoStop [0]{.\EOS\space}%
\providecommand \EOS [0]{\spacefactor3000\relax}%
\providecommand \BibitemShut  [1]{\csname bibitem#1\endcsname}%
\let\auto@bib@innerbib\@empty
\bibitem [{\citenamefont {Bartholomew}\ \emph {et~al.}(1973)\citenamefont
  {Bartholomew}, \citenamefont {Earle}, \citenamefont {A.J.Ferguson},
  \citenamefont {Knowles},\ and\ \citenamefont {Lone}}]{bartholomew}%
  \BibitemOpen
  \bibfield  {author} {\bibinfo {author} {\bibfnamefont {G.~A.}\ \bibnamefont
  {Bartholomew}}, \bibinfo {author} {\bibfnamefont {E.}~\bibnamefont {Earle}},
  \bibinfo {author} {\bibnamefont {A.J.Ferguson}}, \bibinfo {author}
  {\bibfnamefont {J.~W.}\ \bibnamefont {Knowles}}, \ and\ \bibinfo {author}
  {\bibfnamefont {M.~A.}\ \bibnamefont {Lone}},\ }\href {\doibase
  https://doi.org/10.1007/978-1-4615-9044-6_4} {\bibfield  {journal} {\bibinfo
  {journal} {Gamma-Ray Strength Functions. In: Baranger M., Vogt E. (eds)
  Advances in Nuclear Physics. Springer, Boston, MA}\ ,\ \bibinfo {pages}
  {229}} (\bibinfo {year} {1973})}\BibitemShut {NoStop}%
\bibitem [{\citenamefont {Hauser}\ and\ \citenamefont
  {Feshbach}(1952)}]{feshbach}%
  \BibitemOpen
  \bibfield  {author} {\bibinfo {author} {\bibfnamefont {W.}~\bibnamefont
  {Hauser}}\ and\ \bibinfo {author} {\bibfnamefont {H.}~\bibnamefont
  {Feshbach}},\ }\href {\doibase https://doi.org/10.1103/PhysRev.87.366}
  {\bibfield  {journal} {\bibinfo  {journal} {Phys. Rev.}\ }\textbf {\bibinfo
  {volume} {87}},\ \bibinfo {pages} {366} (\bibinfo {year} {1952})}\BibitemShut
  {NoStop}%
\bibitem [{\citenamefont {Arnould}\ \emph {et~al.}(2007)\citenamefont
  {Arnould}, \citenamefont {Goriely},\ and\ \citenamefont
  {Takahashi}}]{arnould}%
  \BibitemOpen
  \bibfield  {author} {\bibinfo {author} {\bibfnamefont {M.}~\bibnamefont
  {Arnould}}, \bibinfo {author} {\bibfnamefont {S.}~\bibnamefont {Goriely}}, \
  and\ \bibinfo {author} {\bibfnamefont {K.}~\bibnamefont {Takahashi}},\ }\href
  {\doibase https://doi.org/10.1016/j.physrep.2007.06.002} {\bibfield
  {journal} {\bibinfo  {journal} {Physics Reports}\ }\textbf {\bibinfo {volume}
  {450}},\ \bibinfo {pages} {97 } (\bibinfo {year} {2007})}\BibitemShut
  {NoStop}%
\bibitem [{\citenamefont {K\"appeler}\ \emph {et~al.}(2011)\citenamefont
  {K\"appeler}, \citenamefont {Gallino}, \citenamefont {Bisterzo},\ and\
  \citenamefont {Aoki}}]{RevModPhys.83.157}%
  \BibitemOpen
  \bibfield  {author} {\bibinfo {author} {\bibfnamefont {F.}~\bibnamefont
  {K\"appeler}}, \bibinfo {author} {\bibfnamefont {R.}~\bibnamefont {Gallino}},
  \bibinfo {author} {\bibfnamefont {S.}~\bibnamefont {Bisterzo}}, \ and\
  \bibinfo {author} {\bibfnamefont {W.}~\bibnamefont {Aoki}},\ }\href {\doibase
  10.1103/RevModPhys.83.157} {\bibfield  {journal} {\bibinfo  {journal} {Rev.
  Mod. Phys.}\ }\textbf {\bibinfo {volume} {83}},\ \bibinfo {pages} {157}
  (\bibinfo {year} {2011})}\BibitemShut {NoStop}%
\bibitem [{\citenamefont {Goriely}(1998)}]{goriely1}%
  \BibitemOpen
  \bibfield  {author} {\bibinfo {author} {\bibfnamefont {S.}~\bibnamefont
  {Goriely}},\ }\href {\doibase https://doi.org/10.1016/S0370-2693(98)00907-1}
  {\bibfield  {journal} {\bibinfo  {journal} {Physics Letters B}\ }\textbf
  {\bibinfo {volume} {436}},\ \bibinfo {pages} {10 } (\bibinfo {year}
  {1998})}\BibitemShut {NoStop}%
\bibitem [{\citenamefont {Bethe}(1936)}]{bethe}%
  \BibitemOpen
  \bibfield  {author} {\bibinfo {author} {\bibfnamefont {H.~A.}\ \bibnamefont
  {Bethe}},\ }\href {\doibase 10.1103/PhysRev.50.332} {\bibfield  {journal}
  {\bibinfo  {journal} {Phys. Rev.}\ }\textbf {\bibinfo {volume} {50}},\
  \bibinfo {pages} {332} (\bibinfo {year} {1936})}\BibitemShut {NoStop}%
\bibitem [{\citenamefont {Decowski}\ \emph {et~al.}(1968)\citenamefont
  {Decowski}, \citenamefont {Grochulski}, \citenamefont {Marcinkowski},
  \citenamefont {Siwek},\ and\ \citenamefont {Wilhelmi}}]{decowski}%
  \BibitemOpen
  \bibfield  {author} {\bibinfo {author} {\bibfnamefont {P.}~\bibnamefont
  {Decowski}}, \bibinfo {author} {\bibfnamefont {W.}~\bibnamefont
  {Grochulski}}, \bibinfo {author} {\bibfnamefont {A.}~\bibnamefont
  {Marcinkowski}}, \bibinfo {author} {\bibfnamefont {K.}~\bibnamefont {Siwek}},
  \ and\ \bibinfo {author} {\bibfnamefont {Z.}~\bibnamefont {Wilhelmi}},\
  }\href {\doibase https://doi.org/10.1016/0375-9474(68)90687-8} {\bibfield
  {journal} {\bibinfo  {journal} {Nuclear Physics A}\ }\textbf {\bibinfo
  {volume} {110}},\ \bibinfo {pages} {129 } (\bibinfo {year}
  {1968})}\BibitemShut {NoStop}%
\bibitem [{\citenamefont {Moretto}(1972)}]{moretto}%
  \BibitemOpen
  \bibfield  {author} {\bibinfo {author} {\bibfnamefont {L.}~\bibnamefont
  {Moretto}},\ }\href {\doibase https://doi.org/10.1016/0375-9474(72)90556-8}
  {\bibfield  {journal} {\bibinfo  {journal} {Nuclear Physics A}\ }\textbf
  {\bibinfo {volume} {185}},\ \bibinfo {pages} {145 } (\bibinfo {year}
  {1972})}\BibitemShut {NoStop}%
\bibitem [{\citenamefont {Hillman}\ and\ \citenamefont
  {Grover}(1969)}]{hillman}%
  \BibitemOpen
  \bibfield  {author} {\bibinfo {author} {\bibfnamefont {M.}~\bibnamefont
  {Hillman}}\ and\ \bibinfo {author} {\bibfnamefont {J.~R.}\ \bibnamefont
  {Grover}},\ }\href {\doibase 10.1103/PhysRev.185.1303} {\bibfield  {journal}
  {\bibinfo  {journal} {Phys. Rev.}\ }\textbf {\bibinfo {volume} {185}},\
  \bibinfo {pages} {1303} (\bibinfo {year} {1969})}\BibitemShut {NoStop}%
\bibitem [{\citenamefont {Hilaire}\ \emph {et~al.}(1998)\citenamefont
  {Hilaire}, \citenamefont {Delaroche},\ and\ \citenamefont
  {Koning}}]{hilaire1}%
  \BibitemOpen
  \bibfield  {author} {\bibinfo {author} {\bibfnamefont {S.}~\bibnamefont
  {Hilaire}}, \bibinfo {author} {\bibfnamefont {J.}~\bibnamefont {Delaroche}},
  \ and\ \bibinfo {author} {\bibfnamefont {A.}~\bibnamefont {Koning}},\ }\href
  {\doibase https://doi.org/10.1016/S0375-9474(98)00003-7} {\bibfield
  {journal} {\bibinfo  {journal} {Nuclear Physics A}\ }\textbf {\bibinfo
  {volume} {632}},\ \bibinfo {pages} {417 } (\bibinfo {year}
  {1998})}\BibitemShut {NoStop}%
\bibitem [{\citenamefont {Hilaire}\ \emph {et~al.}(2001)\citenamefont
  {Hilaire}, \citenamefont {Delaroche},\ and\ \citenamefont
  {Girod}}]{Hilaire2}%
  \BibitemOpen
  \bibfield  {author} {\bibinfo {author} {\bibfnamefont {S.}~\bibnamefont
  {Hilaire}}, \bibinfo {author} {\bibfnamefont {J.}~\bibnamefont {Delaroche}},
  \ and\ \bibinfo {author} {\bibfnamefont {M.}~\bibnamefont {Girod}},\ }\href
  {\doibase 10.1007/s100500170025} {\bibfield  {journal} {\bibinfo  {journal}
  {The European Physical Journal A - Hadrons and Nuclei}\ }\textbf {\bibinfo
  {volume} {12}},\ \bibinfo {pages} {169} (\bibinfo {year} {2001})}\BibitemShut
  {NoStop}%
\bibitem [{\citenamefont {Alhassid}\ \emph {et~al.}(1999)\citenamefont
  {Alhassid}, \citenamefont {Liu},\ and\ \citenamefont {Nakada}}]{alhassid}%
  \BibitemOpen
  \bibfield  {author} {\bibinfo {author} {\bibfnamefont {Y.}~\bibnamefont
  {Alhassid}}, \bibinfo {author} {\bibfnamefont {S.}~\bibnamefont {Liu}}, \
  and\ \bibinfo {author} {\bibfnamefont {H.}~\bibnamefont {Nakada}},\ }\href
  {\doibase 10.1103/PhysRevLett.83.4265} {\bibfield  {journal} {\bibinfo
  {journal} {Phys. Rev. Lett.}\ }\textbf {\bibinfo {volume} {83}},\ \bibinfo
  {pages} {4265} (\bibinfo {year} {1999})}\BibitemShut {NoStop}%
\bibitem [{\citenamefont {Demetriou}\ and\ \citenamefont
  {Goriely}(2001)}]{demetriou}%
  \BibitemOpen
  \bibfield  {author} {\bibinfo {author} {\bibfnamefont {P.}~\bibnamefont
  {Demetriou}}\ and\ \bibinfo {author} {\bibfnamefont {S.}~\bibnamefont
  {Goriely}},\ }\href {\doibase https://doi.org/10.1016/S0375-9474(01)01095-8}
  {\bibfield  {journal} {\bibinfo  {journal} {Nuclear Physics A}\ }\textbf
  {\bibinfo {volume} {695}},\ \bibinfo {pages} {95 } (\bibinfo {year}
  {2001})}\BibitemShut {NoStop}%
\bibitem [{\citenamefont {Hilaire}\ and\ \citenamefont
  {Goriely}(2006)}]{hilaire3}%
  \BibitemOpen
  \bibfield  {author} {\bibinfo {author} {\bibfnamefont {S.}~\bibnamefont
  {Hilaire}}\ and\ \bibinfo {author} {\bibfnamefont {S.}~\bibnamefont
  {Goriely}},\ }\href {\doibase
  https://doi.org/10.1016/j.nuclphysa.2006.08.014} {\bibfield  {journal}
  {\bibinfo  {journal} {Nuclear Physics A}\ }\textbf {\bibinfo {volume}
  {779}},\ \bibinfo {pages} {63 } (\bibinfo {year} {2006})}\BibitemShut
  {NoStop}%
\bibitem [{\citenamefont {Goriely}\ \emph {et~al.}(2008)\citenamefont
  {Goriely}, \citenamefont {Hilaire},\ and\ \citenamefont {Koning}}]{goriely}%
  \BibitemOpen
  \bibfield  {author} {\bibinfo {author} {\bibfnamefont {S.}~\bibnamefont
  {Goriely}}, \bibinfo {author} {\bibfnamefont {S.}~\bibnamefont {Hilaire}}, \
  and\ \bibinfo {author} {\bibfnamefont {A.~J.}\ \bibnamefont {Koning}},\
  }\href {\doibase 10.1103/PhysRevC.78.064307} {\bibfield  {journal} {\bibinfo
  {journal} {Phys. Rev. C}\ }\textbf {\bibinfo {volume} {78}},\ \bibinfo
  {pages} {064307} (\bibinfo {year} {2008})}\BibitemShut {NoStop}%
\bibitem [{\citenamefont {Dietrich}\ and\ \citenamefont
  {Berman}(1988)}]{dietrich}%
  \BibitemOpen
  \bibfield  {author} {\bibinfo {author} {\bibfnamefont {S.~S.}\ \bibnamefont
  {Dietrich}}\ and\ \bibinfo {author} {\bibfnamefont {B.~L.}\ \bibnamefont
  {Berman}},\ }\href {\doibase https://doi.org/10.1016/0092-640X(88)90033-2}
  {\bibfield  {journal} {\bibinfo  {journal} {Atomic Data and Nuclear Data
  Tables}\ }\textbf {\bibinfo {volume} {38}},\ \bibinfo {pages} {199 }
  (\bibinfo {year} {1988})}\BibitemShut {NoStop}%
\bibitem [{\citenamefont {Bracco}\ \emph {et~al.}(2015)\citenamefont {Bracco},
  \citenamefont {Crespi},\ and\ \citenamefont {Lanza}}]{bracco}%
  \BibitemOpen
  \bibfield  {author} {\bibinfo {author} {\bibfnamefont {A.}~\bibnamefont
  {Bracco}}, \bibinfo {author} {\bibfnamefont {F.~C.~L.}\ \bibnamefont
  {Crespi}}, \ and\ \bibinfo {author} {\bibfnamefont {E.~G.}\ \bibnamefont
  {Lanza}},\ }\href {\doibase 10.1140/epja/i2015-15099-6} {\bibfield  {journal}
  {\bibinfo  {journal} {The European Physical Journal A}\ }\textbf {\bibinfo
  {volume} {51}},\ \bibinfo {pages} {99} (\bibinfo {year} {2015})}\BibitemShut
  {NoStop}%
\bibitem [{\citenamefont {Savran}\ \emph {et~al.}(2013)\citenamefont {Savran},
  \citenamefont {Aumann},\ and\ \citenamefont {Zilges}}]{savran}%
  \BibitemOpen
  \bibfield  {author} {\bibinfo {author} {\bibfnamefont {D.}~\bibnamefont
  {Savran}}, \bibinfo {author} {\bibfnamefont {T.}~\bibnamefont {Aumann}}, \
  and\ \bibinfo {author} {\bibfnamefont {A.}~\bibnamefont {Zilges}},\ }\href
  {\doibase https://doi.org/10.1016/j.ppnp.2013.02.003} {\bibfield  {journal}
  {\bibinfo  {journal} {Progress in Particle and Nuclear Physics}\ }\textbf
  {\bibinfo {volume} {70}},\ \bibinfo {pages} {210 } (\bibinfo {year}
  {2013})}\BibitemShut {NoStop}%
\bibitem [{\citenamefont {Schiller}\ \emph {et~al.}(2006)\citenamefont
  {Schiller}, \citenamefont {Voinov}, \citenamefont {Algin}, \citenamefont
  {Becker}, \citenamefont {Bernstein}, \citenamefont {Garrett}, \citenamefont
  {Guttormsen}, \citenamefont {Nelson}, \citenamefont {Rekstad},\ and\
  \citenamefont {Siem}}]{schiller}%
  \BibitemOpen
  \bibfield  {author} {\bibinfo {author} {\bibfnamefont {A.}~\bibnamefont
  {Schiller}}, \bibinfo {author} {\bibfnamefont {A.}~\bibnamefont {Voinov}},
  \bibinfo {author} {\bibfnamefont {E.}~\bibnamefont {Algin}}, \bibinfo
  {author} {\bibfnamefont {J.}~\bibnamefont {Becker}}, \bibinfo {author}
  {\bibfnamefont {L.}~\bibnamefont {Bernstein}}, \bibinfo {author}
  {\bibfnamefont {P.}~\bibnamefont {Garrett}}, \bibinfo {author} {\bibfnamefont
  {M.}~\bibnamefont {Guttormsen}}, \bibinfo {author} {\bibfnamefont
  {R.}~\bibnamefont {Nelson}}, \bibinfo {author} {\bibfnamefont
  {J.}~\bibnamefont {Rekstad}}, \ and\ \bibinfo {author} {\bibfnamefont
  {S.}~\bibnamefont {Siem}},\ }\href {\doibase
  https://doi.org/10.1016/j.physletb.2005.12.043} {\bibfield  {journal}
  {\bibinfo  {journal} {Physics Letters B}\ }\textbf {\bibinfo {volume}
  {633}},\ \bibinfo {pages} {225 } (\bibinfo {year} {2006})}\BibitemShut
  {NoStop}%
\bibitem [{\citenamefont {Heyde}\ \emph {et~al.}(2010)\citenamefont {Heyde},
  \citenamefont {von Neumann-Cosel},\ and\ \citenamefont {Richter}}]{heyde}%
  \BibitemOpen
  \bibfield  {author} {\bibinfo {author} {\bibfnamefont {K.}~\bibnamefont
  {Heyde}}, \bibinfo {author} {\bibfnamefont {P.}~\bibnamefont {von
  Neumann-Cosel}}, \ and\ \bibinfo {author} {\bibfnamefont {A.}~\bibnamefont
  {Richter}},\ }\href {\doibase 10.1103/RevModPhys.82.2365} {\bibfield
  {journal} {\bibinfo  {journal} {Rev. Mod. Phys.}\ }\textbf {\bibinfo {volume}
  {82}},\ \bibinfo {pages} {2365} (\bibinfo {year} {2010})}\BibitemShut
  {NoStop}%
\bibitem [{\citenamefont {Voinov}\ \emph {et~al.}(2004)\citenamefont {Voinov},
  \citenamefont {Algin}, \citenamefont {Agvaanluvsan}, \citenamefont {Belgya},
  \citenamefont {Chankova}, \citenamefont {Guttormsen}, \citenamefont
  {Mitchell}, \citenamefont {Rekstad}, \citenamefont {Schiller},\ and\
  \citenamefont {Siem}}]{voivnov}%
  \BibitemOpen
  \bibfield  {author} {\bibinfo {author} {\bibfnamefont {A.}~\bibnamefont
  {Voinov}}, \bibinfo {author} {\bibfnamefont {E.}~\bibnamefont {Algin}},
  \bibinfo {author} {\bibfnamefont {U.}~\bibnamefont {Agvaanluvsan}}, \bibinfo
  {author} {\bibfnamefont {T.}~\bibnamefont {Belgya}}, \bibinfo {author}
  {\bibfnamefont {R.}~\bibnamefont {Chankova}}, \bibinfo {author}
  {\bibfnamefont {M.}~\bibnamefont {Guttormsen}}, \bibinfo {author}
  {\bibfnamefont {G.~E.}\ \bibnamefont {Mitchell}}, \bibinfo {author}
  {\bibfnamefont {J.}~\bibnamefont {Rekstad}}, \bibinfo {author} {\bibfnamefont
  {A.}~\bibnamefont {Schiller}}, \ and\ \bibinfo {author} {\bibfnamefont
  {S.}~\bibnamefont {Siem}},\ }\href {\doibase 10.1103/PhysRevLett.93.142504}
  {\bibfield  {journal} {\bibinfo  {journal} {Phys. Rev. Lett.}\ }\textbf
  {\bibinfo {volume} {93}},\ \bibinfo {pages} {142504} (\bibinfo {year}
  {2004})}\BibitemShut {NoStop}%
\bibitem [{\citenamefont {Guttormsen}\ \emph {et~al.}(2005)\citenamefont
  {Guttormsen}, \citenamefont {Chankova}, \citenamefont {Agvaanluvsan},
  \citenamefont {Algin}, \citenamefont {Bernstein}, \citenamefont
  {Ingebretsen}, \citenamefont {L\"onnroth}, \citenamefont {Messelt},
  \citenamefont {Mitchell}, \citenamefont {Rekstad}, \citenamefont {Schiller},
  \citenamefont {Siem}, \citenamefont {Sunde}, \citenamefont {Voinov},\ and\
  \citenamefont {\O{}deg\aa{}rd}}]{guttormsen}%
  \BibitemOpen
  \bibfield  {author} {\bibinfo {author} {\bibfnamefont {M.}~\bibnamefont
  {Guttormsen}}, \bibinfo {author} {\bibfnamefont {R.}~\bibnamefont
  {Chankova}}, \bibinfo {author} {\bibfnamefont {U.}~\bibnamefont
  {Agvaanluvsan}}, \bibinfo {author} {\bibfnamefont {E.}~\bibnamefont {Algin}},
  \bibinfo {author} {\bibfnamefont {L.~A.}\ \bibnamefont {Bernstein}}, \bibinfo
  {author} {\bibfnamefont {F.}~\bibnamefont {Ingebretsen}}, \bibinfo {author}
  {\bibfnamefont {T.}~\bibnamefont {L\"onnroth}}, \bibinfo {author}
  {\bibfnamefont {S.}~\bibnamefont {Messelt}}, \bibinfo {author} {\bibfnamefont
  {G.~E.}\ \bibnamefont {Mitchell}}, \bibinfo {author} {\bibfnamefont
  {J.}~\bibnamefont {Rekstad}}, \bibinfo {author} {\bibfnamefont
  {A.}~\bibnamefont {Schiller}}, \bibinfo {author} {\bibfnamefont
  {S.}~\bibnamefont {Siem}}, \bibinfo {author} {\bibfnamefont {A.~C.}\
  \bibnamefont {Sunde}}, \bibinfo {author} {\bibfnamefont {A.}~\bibnamefont
  {Voinov}}, \ and\ \bibinfo {author} {\bibfnamefont {S.}~\bibnamefont
  {\O{}deg\aa{}rd}},\ }\href {\doibase 10.1103/PhysRevC.71.044307} {\bibfield
  {journal} {\bibinfo  {journal} {Phys. Rev. C}\ }\textbf {\bibinfo {volume}
  {71}},\ \bibinfo {pages} {044307} (\bibinfo {year} {2005})}\BibitemShut
  {NoStop}%
\bibitem [{\citenamefont {Wiedeking}\ \emph {et~al.}(2012)\citenamefont
  {Wiedeking}, \citenamefont {Bernstein}, \citenamefont
  {Krti\ifmmode~\check{c}\else \v{c}\fi{}ka}, \citenamefont {Bleuel},
  \citenamefont {Allmond}, \citenamefont {Basunia}, \citenamefont {Burke},
  \citenamefont {Fallon}, \citenamefont {Firestone}, \citenamefont {Goldblum},
  \citenamefont {Hatarik}, \citenamefont {Lake}, \citenamefont {Lee},
  \citenamefont {Lesher}, \citenamefont {Paschalis}, \citenamefont {Petri},
  \citenamefont {Phair},\ and\ \citenamefont {Scielzo}}]{wiedeking}%
  \BibitemOpen
  \bibfield  {author} {\bibinfo {author} {\bibfnamefont {M.}~\bibnamefont
  {Wiedeking}}, \bibinfo {author} {\bibfnamefont {L.~A.}\ \bibnamefont
  {Bernstein}}, \bibinfo {author} {\bibfnamefont {M.}~\bibnamefont
  {Krti\ifmmode~\check{c}\else \v{c}\fi{}ka}}, \bibinfo {author} {\bibfnamefont
  {D.~L.}\ \bibnamefont {Bleuel}}, \bibinfo {author} {\bibfnamefont {J.~M.}\
  \bibnamefont {Allmond}}, \bibinfo {author} {\bibfnamefont {M.~S.}\
  \bibnamefont {Basunia}}, \bibinfo {author} {\bibfnamefont {J.~T.}\
  \bibnamefont {Burke}}, \bibinfo {author} {\bibfnamefont {P.}~\bibnamefont
  {Fallon}}, \bibinfo {author} {\bibfnamefont {R.~B.}\ \bibnamefont
  {Firestone}}, \bibinfo {author} {\bibfnamefont {B.~L.}\ \bibnamefont
  {Goldblum}}, \bibinfo {author} {\bibfnamefont {R.}~\bibnamefont {Hatarik}},
  \bibinfo {author} {\bibfnamefont {P.~T.}\ \bibnamefont {Lake}}, \bibinfo
  {author} {\bibfnamefont {I.-Y.}\ \bibnamefont {Lee}}, \bibinfo {author}
  {\bibfnamefont {S.~R.}\ \bibnamefont {Lesher}}, \bibinfo {author}
  {\bibfnamefont {S.}~\bibnamefont {Paschalis}}, \bibinfo {author}
  {\bibfnamefont {M.}~\bibnamefont {Petri}}, \bibinfo {author} {\bibfnamefont
  {L.}~\bibnamefont {Phair}}, \ and\ \bibinfo {author} {\bibfnamefont {N.~D.}\
  \bibnamefont {Scielzo}},\ }\href {\doibase 10.1103/PhysRevLett.108.162503}
  {\bibfield  {journal} {\bibinfo  {journal} {Phys. Rev. Lett.}\ }\textbf
  {\bibinfo {volume} {108}},\ \bibinfo {pages} {162503} (\bibinfo {year}
  {2012})}\BibitemShut {NoStop}%
\bibitem [{\citenamefont {Larsen}\ \emph {et~al.}(2013)\citenamefont {Larsen},
  \citenamefont {Blasi}, \citenamefont {Bracco}, \citenamefont {Camera},
  \citenamefont {Eriksen}, \citenamefont {G\"orgen}, \citenamefont
  {Guttormsen}, \citenamefont {Hagen}, \citenamefont {Leoni}, \citenamefont
  {Million}, \citenamefont {Nyhus}, \citenamefont {Renstr\o{}m}, \citenamefont
  {Rose}, \citenamefont {Ruud}, \citenamefont {Siem}, \citenamefont {Tornyi},
  \citenamefont {Tveten}, \citenamefont {Voinov},\ and\ \citenamefont
  {Wiedeking}}]{larsen}%
  \BibitemOpen
  \bibfield  {author} {\bibinfo {author} {\bibfnamefont {A.~C.}\ \bibnamefont
  {Larsen}}, \bibinfo {author} {\bibfnamefont {N.}~\bibnamefont {Blasi}},
  \bibinfo {author} {\bibfnamefont {A.}~\bibnamefont {Bracco}}, \bibinfo
  {author} {\bibfnamefont {F.}~\bibnamefont {Camera}}, \bibinfo {author}
  {\bibfnamefont {T.~K.}\ \bibnamefont {Eriksen}}, \bibinfo {author}
  {\bibfnamefont {A.}~\bibnamefont {G\"orgen}}, \bibinfo {author}
  {\bibfnamefont {M.}~\bibnamefont {Guttormsen}}, \bibinfo {author}
  {\bibfnamefont {T.~W.}\ \bibnamefont {Hagen}}, \bibinfo {author}
  {\bibfnamefont {S.}~\bibnamefont {Leoni}}, \bibinfo {author} {\bibfnamefont
  {B.}~\bibnamefont {Million}}, \bibinfo {author} {\bibfnamefont {H.~T.}\
  \bibnamefont {Nyhus}}, \bibinfo {author} {\bibfnamefont {T.}~\bibnamefont
  {Renstr\o{}m}}, \bibinfo {author} {\bibfnamefont {S.~J.}\ \bibnamefont
  {Rose}}, \bibinfo {author} {\bibfnamefont {I.~E.}\ \bibnamefont {Ruud}},
  \bibinfo {author} {\bibfnamefont {S.}~\bibnamefont {Siem}}, \bibinfo {author}
  {\bibfnamefont {T.}~\bibnamefont {Tornyi}}, \bibinfo {author} {\bibfnamefont
  {G.~M.}\ \bibnamefont {Tveten}}, \bibinfo {author} {\bibfnamefont {A.~V.}\
  \bibnamefont {Voinov}}, \ and\ \bibinfo {author} {\bibfnamefont
  {M.}~\bibnamefont {Wiedeking}},\ }\href {\doibase
  10.1103/PhysRevLett.111.242504} {\bibfield  {journal} {\bibinfo  {journal}
  {Phys. Rev. Lett.}\ }\textbf {\bibinfo {volume} {111}},\ \bibinfo {pages}
  {242504} (\bibinfo {year} {2013})}\BibitemShut {NoStop}%
\bibitem [{\citenamefont {Jones}\ \emph {et~al.}(2018)\citenamefont {Jones},
  \citenamefont {Macchiavelli}, \citenamefont {Wiedeking}, \citenamefont
  {Bernstein}, \citenamefont {Crawford}, \citenamefont {Campbell},
  \citenamefont {Clark}, \citenamefont {Cromaz}, \citenamefont {Fallon},
  \citenamefont {Lee}, \citenamefont {Salathe}, \citenamefont {Wiens},
  \citenamefont {Ayangeakaa}, \citenamefont {Bleuel}, \citenamefont {Bottoni},
  \citenamefont {Carpenter}, \citenamefont {Davids}, \citenamefont {Elson},
  \citenamefont {G\"orgen}, \citenamefont {Guttormsen}, \citenamefont
  {Janssens}, \citenamefont {Kinnison}, \citenamefont {Kirsch}, \citenamefont
  {Larsen}, \citenamefont {Lauritsen}, \citenamefont {Reviol}, \citenamefont
  {Sarantites}, \citenamefont {Siem}, \citenamefont {Voinov},\ and\
  \citenamefont {Zhu}}]{jones}%
  \BibitemOpen
  \bibfield  {author} {\bibinfo {author} {\bibfnamefont {M.~D.}\ \bibnamefont
  {Jones}}, \bibinfo {author} {\bibfnamefont {A.~O.}\ \bibnamefont
  {Macchiavelli}}, \bibinfo {author} {\bibfnamefont {M.}~\bibnamefont
  {Wiedeking}}, \bibinfo {author} {\bibfnamefont {L.~A.}\ \bibnamefont
  {Bernstein}}, \bibinfo {author} {\bibfnamefont {H.~L.}\ \bibnamefont
  {Crawford}}, \bibinfo {author} {\bibfnamefont {C.~M.}\ \bibnamefont
  {Campbell}}, \bibinfo {author} {\bibfnamefont {R.~M.}\ \bibnamefont {Clark}},
  \bibinfo {author} {\bibfnamefont {M.}~\bibnamefont {Cromaz}}, \bibinfo
  {author} {\bibfnamefont {P.}~\bibnamefont {Fallon}}, \bibinfo {author}
  {\bibfnamefont {I.~Y.}\ \bibnamefont {Lee}}, \bibinfo {author} {\bibfnamefont
  {M.}~\bibnamefont {Salathe}}, \bibinfo {author} {\bibfnamefont
  {A.}~\bibnamefont {Wiens}}, \bibinfo {author} {\bibfnamefont {A.~D.}\
  \bibnamefont {Ayangeakaa}}, \bibinfo {author} {\bibfnamefont {D.~L.}\
  \bibnamefont {Bleuel}}, \bibinfo {author} {\bibfnamefont {S.}~\bibnamefont
  {Bottoni}}, \bibinfo {author} {\bibfnamefont {M.~P.}\ \bibnamefont
  {Carpenter}}, \bibinfo {author} {\bibfnamefont {H.~M.}\ \bibnamefont
  {Davids}}, \bibinfo {author} {\bibfnamefont {J.}~\bibnamefont {Elson}},
  \bibinfo {author} {\bibfnamefont {A.}~\bibnamefont {G\"orgen}}, \bibinfo
  {author} {\bibfnamefont {M.}~\bibnamefont {Guttormsen}}, \bibinfo {author}
  {\bibfnamefont {R.~V.~F.}\ \bibnamefont {Janssens}}, \bibinfo {author}
  {\bibfnamefont {J.~E.}\ \bibnamefont {Kinnison}}, \bibinfo {author}
  {\bibfnamefont {L.}~\bibnamefont {Kirsch}}, \bibinfo {author} {\bibfnamefont
  {A.~C.}\ \bibnamefont {Larsen}}, \bibinfo {author} {\bibfnamefont
  {T.}~\bibnamefont {Lauritsen}}, \bibinfo {author} {\bibfnamefont
  {W.}~\bibnamefont {Reviol}}, \bibinfo {author} {\bibfnamefont {D.~G.}\
  \bibnamefont {Sarantites}}, \bibinfo {author} {\bibfnamefont
  {S.}~\bibnamefont {Siem}}, \bibinfo {author} {\bibfnamefont {A.~V.}\
  \bibnamefont {Voinov}}, \ and\ \bibinfo {author} {\bibfnamefont
  {S.}~\bibnamefont {Zhu}},\ }\href {\doibase 10.1103/PhysRevC.97.024327}
  {\bibfield  {journal} {\bibinfo  {journal} {Phys. Rev. C}\ }\textbf {\bibinfo
  {volume} {97}},\ \bibinfo {pages} {024327} (\bibinfo {year}
  {2018})}\BibitemShut {NoStop}%
\bibitem [{\citenamefont {Goriely}\ \emph {et~al.}(2018)\citenamefont
  {Goriely}, \citenamefont {Hilaire}, \citenamefont {P\'eru},\ and\
  \citenamefont {Sieja}}]{Gor18}%
  \BibitemOpen
  \bibfield  {author} {\bibinfo {author} {\bibfnamefont {S.}~\bibnamefont
  {Goriely}}, \bibinfo {author} {\bibfnamefont {S.}~\bibnamefont {Hilaire}},
  \bibinfo {author} {\bibfnamefont {S.}~\bibnamefont {P\'eru}}, \ and\ \bibinfo
  {author} {\bibfnamefont {K.}~\bibnamefont {Sieja}},\ }\href {\doibase
  10.1103/PhysRevC.98.014327} {\bibfield  {journal} {\bibinfo  {journal} {Phys.
  Rev. C}\ }\textbf {\bibinfo {volume} {98}},\ \bibinfo {pages} {014327}
  (\bibinfo {year} {2018})}\BibitemShut {NoStop}%
\bibitem [{\citenamefont {Schwengner}\ \emph {et~al.}(2013)\citenamefont
  {Schwengner}, \citenamefont {Frauendorf},\ and\ \citenamefont
  {Larsen}}]{schwengner}%
  \BibitemOpen
  \bibfield  {author} {\bibinfo {author} {\bibfnamefont {R.}~\bibnamefont
  {Schwengner}}, \bibinfo {author} {\bibfnamefont {S.}~\bibnamefont
  {Frauendorf}}, \ and\ \bibinfo {author} {\bibfnamefont {A.~C.}\ \bibnamefont
  {Larsen}},\ }\href {\doibase 10.1103/PhysRevLett.111.232504} {\bibfield
  {journal} {\bibinfo  {journal} {Phys. Rev. Lett.}\ }\textbf {\bibinfo
  {volume} {111}},\ \bibinfo {pages} {232504} (\bibinfo {year}
  {2013})}\BibitemShut {NoStop}%
\bibitem [{\citenamefont {Brown}\ and\ \citenamefont {Larsen}(2014)}]{brown}%
  \BibitemOpen
  \bibfield  {author} {\bibinfo {author} {\bibfnamefont {B.~A.}\ \bibnamefont
  {Brown}}\ and\ \bibinfo {author} {\bibfnamefont {A.~C.}\ \bibnamefont
  {Larsen}},\ }\href {\doibase 10.1103/PhysRevLett.113.252502} {\bibfield
  {journal} {\bibinfo  {journal} {Phys. Rev. Lett.}\ }\textbf {\bibinfo
  {volume} {113}},\ \bibinfo {pages} {252502} (\bibinfo {year}
  {2014})}\BibitemShut {NoStop}%
\bibitem [{\citenamefont {Sieja}(2017)}]{sieja}%
  \BibitemOpen
  \bibfield  {author} {\bibinfo {author} {\bibfnamefont {K.}~\bibnamefont
  {Sieja}},\ }\href {\doibase 10.1103/PhysRevLett.119.052502} {\bibfield
  {journal} {\bibinfo  {journal} {Phys. Rev. Lett.}\ }\textbf {\bibinfo
  {volume} {119}},\ \bibinfo {pages} {052502} (\bibinfo {year}
  {2017})}\BibitemShut {NoStop}%
\bibitem [{\citenamefont {Schwengner}\ \emph {et~al.}(2017)\citenamefont
  {Schwengner}, \citenamefont {Frauendorf},\ and\ \citenamefont
  {Brown}}]{schwengner2}%
  \BibitemOpen
  \bibfield  {author} {\bibinfo {author} {\bibfnamefont {R.}~\bibnamefont
  {Schwengner}}, \bibinfo {author} {\bibfnamefont {S.}~\bibnamefont
  {Frauendorf}}, \ and\ \bibinfo {author} {\bibfnamefont {B.~A.}\ \bibnamefont
  {Brown}},\ }\href {\doibase 10.1103/PhysRevLett.118.092502} {\bibfield
  {journal} {\bibinfo  {journal} {Phys. Rev. Lett.}\ }\textbf {\bibinfo
  {volume} {118}},\ \bibinfo {pages} {092502} (\bibinfo {year}
  {2017})}\BibitemShut {NoStop}%
\bibitem [{\citenamefont {Simon}\ \emph {et~al.}(2016)\citenamefont {Simon},
  \citenamefont {Guttormsen}, \citenamefont {Larsen}, \citenamefont {Beausang},
  \citenamefont {Humby}, \citenamefont {Burke}, \citenamefont {Casperson},
  \citenamefont {Hughes}, \citenamefont {Ross}, \citenamefont {Allmond},
  \citenamefont {Chyzh}, \citenamefont {Dag}, \citenamefont {Koglin},
  \citenamefont {McCleskey}, \citenamefont {McCleskey}, \citenamefont {Ota},\
  and\ \citenamefont {Saastamoinen}}]{simon16}%
  \BibitemOpen
  \bibfield  {author} {\bibinfo {author} {\bibfnamefont {A.}~\bibnamefont
  {Simon}}, \bibinfo {author} {\bibfnamefont {M.}~\bibnamefont {Guttormsen}},
  \bibinfo {author} {\bibfnamefont {A.~C.}\ \bibnamefont {Larsen}}, \bibinfo
  {author} {\bibfnamefont {C.~W.}\ \bibnamefont {Beausang}}, \bibinfo {author}
  {\bibfnamefont {P.}~\bibnamefont {Humby}}, \bibinfo {author} {\bibfnamefont
  {J.~T.}\ \bibnamefont {Burke}}, \bibinfo {author} {\bibfnamefont {R.~J.}\
  \bibnamefont {Casperson}}, \bibinfo {author} {\bibfnamefont {R.~O.}\
  \bibnamefont {Hughes}}, \bibinfo {author} {\bibfnamefont {T.~J.}\
  \bibnamefont {Ross}}, \bibinfo {author} {\bibfnamefont {J.~M.}\ \bibnamefont
  {Allmond}}, \bibinfo {author} {\bibfnamefont {R.}~\bibnamefont {Chyzh}},
  \bibinfo {author} {\bibfnamefont {M.}~\bibnamefont {Dag}}, \bibinfo {author}
  {\bibfnamefont {J.}~\bibnamefont {Koglin}}, \bibinfo {author} {\bibfnamefont
  {E.}~\bibnamefont {McCleskey}}, \bibinfo {author} {\bibfnamefont
  {M.}~\bibnamefont {McCleskey}}, \bibinfo {author} {\bibfnamefont
  {S.}~\bibnamefont {Ota}}, \ and\ \bibinfo {author} {\bibfnamefont
  {A.}~\bibnamefont {Saastamoinen}},\ }\href {\doibase
  10.1103/PhysRevC.93.034303} {\bibfield  {journal} {\bibinfo  {journal} {Phys.
  Rev. C}\ }\textbf {\bibinfo {volume} {93}},\ \bibinfo {pages} {034303}
  (\bibinfo {year} {2016})}\BibitemShut {NoStop}%
\bibitem [{\citenamefont {Hughes}\ \emph {et~al.}(2017)\citenamefont {Hughes},
  \citenamefont {Burke}, \citenamefont {Casperson}, \citenamefont {Ota},
  \citenamefont {Fisher}, \citenamefont {Parker}, \citenamefont {Beausang},
  \citenamefont {Dag}, \citenamefont {Humby}, \citenamefont {Koglin},
  \citenamefont {McCleskey}, \citenamefont {McIntosh}, \citenamefont
  {Saastamoinen}, \citenamefont {Tamashiro}, \citenamefont {Wilson},\ and\
  \citenamefont {Wu}}]{hughes17}%
  \BibitemOpen
  \bibfield  {author} {\bibinfo {author} {\bibfnamefont {R.}~\bibnamefont
  {Hughes}}, \bibinfo {author} {\bibfnamefont {J.}~\bibnamefont {Burke}},
  \bibinfo {author} {\bibfnamefont {R.}~\bibnamefont {Casperson}}, \bibinfo
  {author} {\bibfnamefont {S.}~\bibnamefont {Ota}}, \bibinfo {author}
  {\bibfnamefont {S.}~\bibnamefont {Fisher}}, \bibinfo {author} {\bibfnamefont
  {J.}~\bibnamefont {Parker}}, \bibinfo {author} {\bibfnamefont
  {C.}~\bibnamefont {Beausang}}, \bibinfo {author} {\bibfnamefont
  {M.}~\bibnamefont {Dag}}, \bibinfo {author} {\bibfnamefont {P.}~\bibnamefont
  {Humby}}, \bibinfo {author} {\bibfnamefont {J.}~\bibnamefont {Koglin}},
  \bibinfo {author} {\bibfnamefont {E.}~\bibnamefont {McCleskey}}, \bibinfo
  {author} {\bibfnamefont {A.}~\bibnamefont {McIntosh}}, \bibinfo {author}
  {\bibfnamefont {A.}~\bibnamefont {Saastamoinen}}, \bibinfo {author}
  {\bibfnamefont {A.}~\bibnamefont {Tamashiro}}, \bibinfo {author}
  {\bibfnamefont {E.}~\bibnamefont {Wilson}}, \ and\ \bibinfo {author}
  {\bibfnamefont {T.}~\bibnamefont {Wu}},\ }\href {\doibase
  https://doi.org/10.1016/j.nima.2017.03.012} {\bibfield  {journal} {\bibinfo
  {journal} {Nucl. Instrum. Methods Phys. Res. A}\ }\textbf {\bibinfo {volume}
  {856}},\ \bibinfo {pages} {47} (\bibinfo {year} {2017})}\BibitemShut
  {NoStop}%
\bibitem [{\citenamefont {Larsen}\ \emph
  {et~al.}(2011{\natexlab{a}})\citenamefont {Larsen}, \citenamefont
  {Guttormsen}, \citenamefont {Krti\ifmmode~\check{c}\else \v{c}\fi{}ka},
  \citenamefont {B\ifmmode~\check{e}\else \v{e}\fi{}t\'ak}, \citenamefont
  {B\"urger}, \citenamefont {G\"orgen}, \citenamefont {Nyhus}, \citenamefont
  {Rekstad}, \citenamefont {Schiller}, \citenamefont {Siem}, \citenamefont
  {Toft}, \citenamefont {Tveten}, \citenamefont {Voinov},\ and\ \citenamefont
  {Wikan}}]{oslo_method}%
  \BibitemOpen
  \bibfield  {author} {\bibinfo {author} {\bibfnamefont {A.~C.}\ \bibnamefont
  {Larsen}}, \bibinfo {author} {\bibfnamefont {M.}~\bibnamefont {Guttormsen}},
  \bibinfo {author} {\bibfnamefont {M.}~\bibnamefont
  {Krti\ifmmode~\check{c}\else \v{c}\fi{}ka}}, \bibinfo {author} {\bibfnamefont
  {E.}~\bibnamefont {B\ifmmode~\check{e}\else \v{e}\fi{}t\'ak}}, \bibinfo
  {author} {\bibfnamefont {A.}~\bibnamefont {B\"urger}}, \bibinfo {author}
  {\bibfnamefont {A.}~\bibnamefont {G\"orgen}}, \bibinfo {author}
  {\bibfnamefont {H.~T.}\ \bibnamefont {Nyhus}}, \bibinfo {author}
  {\bibfnamefont {J.}~\bibnamefont {Rekstad}}, \bibinfo {author} {\bibfnamefont
  {A.}~\bibnamefont {Schiller}}, \bibinfo {author} {\bibfnamefont
  {S.}~\bibnamefont {Siem}}, \bibinfo {author} {\bibfnamefont {H.~K.}\
  \bibnamefont {Toft}}, \bibinfo {author} {\bibfnamefont {G.~M.}\ \bibnamefont
  {Tveten}}, \bibinfo {author} {\bibfnamefont {A.~V.}\ \bibnamefont {Voinov}},
  \ and\ \bibinfo {author} {\bibfnamefont {K.}~\bibnamefont {Wikan}},\ }\href
  {\doibase 10.1103/PhysRevC.83.034315} {\bibfield  {journal} {\bibinfo
  {journal} {Phys. Rev. C}\ }\textbf {\bibinfo {volume} {83}},\ \bibinfo
  {pages} {034315} (\bibinfo {year} {2011}{\natexlab{a}})}\BibitemShut
  {NoStop}%
\bibitem [{\citenamefont {Agostinelli~{\it{et al.}}}(2003)}]{geant4}%
  \BibitemOpen
  \bibfield  {author} {\bibinfo {author} {\bibfnamefont {S.}~\bibnamefont
  {Agostinelli~{\it{et al.}}}},\ }\href {\doibase
  https://doi.org/10.1016/S0168-9002(03)01368-8} {\bibfield  {journal}
  {\bibinfo  {journal} {Nuclear Instruments and Methods in Physics Research
  Section A: Accelerators, Spectrometers, Detectors and Associated Equipment}\
  }\textbf {\bibinfo {volume} {506}},\ \bibinfo {pages} {250 } (\bibinfo {year}
  {2003})}\BibitemShut {NoStop}%
\bibitem [{\citenamefont {Guttormsen}\ \emph {et~al.}(1996)\citenamefont
  {Guttormsen}, \citenamefont {Tveter}, \citenamefont {Bergholt}, \citenamefont
  {Ingebretsen},\ and\ \citenamefont {Rekstad}}]{unfolding}%
  \BibitemOpen
  \bibfield  {author} {\bibinfo {author} {\bibfnamefont {M.}~\bibnamefont
  {Guttormsen}}, \bibinfo {author} {\bibfnamefont {T.}~\bibnamefont {Tveter}},
  \bibinfo {author} {\bibfnamefont {L.}~\bibnamefont {Bergholt}}, \bibinfo
  {author} {\bibfnamefont {F.}~\bibnamefont {Ingebretsen}}, \ and\ \bibinfo
  {author} {\bibfnamefont {J.}~\bibnamefont {Rekstad}},\ }\href {\doibase
  https://doi.org/10.1016/0168-9002(96)00197-0} {\bibfield  {journal} {\bibinfo
   {journal} {Nuclear Instruments and Methods in Physics Research Section A:
  Accelerators, Spectrometers, Detectors and Associated Equipment}\ }\textbf
  {\bibinfo {volume} {374}},\ \bibinfo {pages} {371 } (\bibinfo {year}
  {1996})}\BibitemShut {NoStop}%
\bibitem [{\citenamefont {Larsen}\ \emph
  {et~al.}(2011{\natexlab{b}})\citenamefont {Larsen}, \citenamefont
  {Guttormsen}, \citenamefont {Krti\ifmmode~\check{c}\else \v{c}\fi{}ka},
  \citenamefont {B\ifmmode~\check{e}\else \v{e}\fi{}t\'ak}, \citenamefont
  {B\"urger}, \citenamefont {G\"orgen}, \citenamefont {Nyhus}, \citenamefont
  {Rekstad}, \citenamefont {Schiller}, \citenamefont {Siem}, \citenamefont
  {Toft}, \citenamefont {Tveten}, \citenamefont {Voinov},\ and\ \citenamefont
  {Wikan}}]{first-generation2}%
  \BibitemOpen
  \bibfield  {author} {\bibinfo {author} {\bibfnamefont {A.~C.}\ \bibnamefont
  {Larsen}}, \bibinfo {author} {\bibfnamefont {M.}~\bibnamefont {Guttormsen}},
  \bibinfo {author} {\bibfnamefont {M.}~\bibnamefont
  {Krti\ifmmode~\check{c}\else \v{c}\fi{}ka}}, \bibinfo {author} {\bibfnamefont
  {E.}~\bibnamefont {B\ifmmode~\check{e}\else \v{e}\fi{}t\'ak}}, \bibinfo
  {author} {\bibfnamefont {A.}~\bibnamefont {B\"urger}}, \bibinfo {author}
  {\bibfnamefont {A.}~\bibnamefont {G\"orgen}}, \bibinfo {author}
  {\bibfnamefont {H.~T.}\ \bibnamefont {Nyhus}}, \bibinfo {author}
  {\bibfnamefont {J.}~\bibnamefont {Rekstad}}, \bibinfo {author} {\bibfnamefont
  {A.}~\bibnamefont {Schiller}}, \bibinfo {author} {\bibfnamefont
  {S.}~\bibnamefont {Siem}}, \bibinfo {author} {\bibfnamefont {H.~K.}\
  \bibnamefont {Toft}}, \bibinfo {author} {\bibfnamefont {G.~M.}\ \bibnamefont
  {Tveten}}, \bibinfo {author} {\bibfnamefont {A.~V.}\ \bibnamefont {Voinov}},
  \ and\ \bibinfo {author} {\bibfnamefont {K.}~\bibnamefont {Wikan}},\ }\href
  {\doibase 10.1103/PhysRevC.83.034315} {\bibfield  {journal} {\bibinfo
  {journal} {Phys. Rev. C}\ }\textbf {\bibinfo {volume} {83}},\ \bibinfo
  {pages} {034315} (\bibinfo {year} {2011}{\natexlab{b}})}\BibitemShut
  {NoStop}%
\bibitem [{\citenamefont {Guttormsen}\ \emph {et~al.}(1987)\citenamefont
  {Guttormsen}, \citenamefont {RamsÃ¸y},\ and\ \citenamefont
  {Rekstad}}]{first-generation}%
  \BibitemOpen
  \bibfield  {author} {\bibinfo {author} {\bibfnamefont {M.}~\bibnamefont
  {Guttormsen}}, \bibinfo {author} {\bibfnamefont {T.}~\bibnamefont
  {RamsÃ¸y}}, \ and\ \bibinfo {author} {\bibfnamefont {J.}~\bibnamefont
  {Rekstad}},\ }\href {\doibase https://doi.org/10.1016/0168-9002(87)91221-6}
  {\bibfield  {journal} {\bibinfo  {journal} {Nuclear Instruments and Methods
  in Physics Research Section A: Accelerators, Spectrometers, Detectors and
  Associated Equipment}\ }\textbf {\bibinfo {volume} {255}},\ \bibinfo {pages}
  {518 } (\bibinfo {year} {1987})}\BibitemShut {NoStop}%
\bibitem [{\citenamefont {Mughabghab}(2018)}]{mughabghab}%
  \BibitemOpen
  \bibfield  {author} {\bibinfo {author} {\bibfnamefont {S.}~\bibnamefont
  {Mughabghab}},\ }in\ \href {\doibase
  https://doi.org/10.1016/B978-0-44-463780-2.00015-3} {\emph {\bibinfo
  {booktitle} {Atlas of Neutron Resonances (Sixth Edition)}}},\ \bibinfo
  {editor} {edited by\ \bibinfo {editor} {\bibfnamefont {S.}~\bibnamefont
  {Mughabghab}}}\ (\bibinfo  {publisher} {Elsevier},\ \bibinfo {address}
  {Amsterdam},\ \bibinfo {year} {2018})\ \bibinfo {edition} {sixth edition}\
  ed.,\ pp.\ \bibinfo {pages} {111 -- 679}\BibitemShut {NoStop}%
\bibitem [{\citenamefont {Brink}(1955)}]{brink}%
  \BibitemOpen
  \bibfield  {author} {\bibinfo {author} {\bibfnamefont {D.~M.}\ \bibnamefont
  {Brink}},\ }\href@noop {} {\bibfield  {journal} {\bibinfo  {journal} {Ph.D.
  thesis, Oxford University}\ } (\bibinfo {year} {1955})}\BibitemShut {NoStop}%
\bibitem [{\citenamefont {Axel}(1962)}]{axel}%
  \BibitemOpen
  \bibfield  {author} {\bibinfo {author} {\bibfnamefont {P.}~\bibnamefont
  {Axel}},\ }\href {\doibase 10.1103/PhysRev.126.671} {\bibfield  {journal}
  {\bibinfo  {journal} {Phys. Rev.}\ }\textbf {\bibinfo {volume} {126}},\
  \bibinfo {pages} {671} (\bibinfo {year} {1962})}\BibitemShut {NoStop}%
\bibitem [{nnd(2019)}]{nndc}%
  \BibitemOpen
  \href@noop {} {\bibfield  {journal} {\bibinfo  {journal} {Available online at
  https://www.nndc.bnl.gov/nudat2/}\ } (\bibinfo {year} {2019})}\BibitemShut
  {NoStop}%
\bibitem [{\citenamefont {von Egidy}\ and\ \citenamefont
  {Bucurescu}(2005)}]{egidy}%
  \BibitemOpen
  \bibfield  {author} {\bibinfo {author} {\bibfnamefont {T.}~\bibnamefont {von
  Egidy}}\ and\ \bibinfo {author} {\bibfnamefont {D.}~\bibnamefont
  {Bucurescu}},\ }\href {\doibase 10.1103/PhysRevC.72.044311} {\bibfield
  {journal} {\bibinfo  {journal} {Phys. Rev. C}\ }\textbf {\bibinfo {volume}
  {72}},\ \bibinfo {pages} {044311} (\bibinfo {year} {2005})}\BibitemShut
  {NoStop}%
\bibitem [{\citenamefont {Renstr\o{}m}\ \emph {et~al.}()\citenamefont
  {Renstr\o{}m}, \citenamefont {Utsunomiya}, \citenamefont {Nyhus},
  \citenamefont {Larsen}, \citenamefont {Guttormsen}, \citenamefont {Tveten},
  \citenamefont {Filipescu}, \citenamefont {Gheorghe}, \citenamefont {Goriely},
  \citenamefont {Hilaire}, \citenamefont {Lui}, \citenamefont {Midtb\o{}},
  \citenamefont {Peru}, \citenamefont {Shima}, \citenamefont {Siem},\ and\
  \citenamefont {Tesileanu}}]{Dy}%
  \BibitemOpen
  \bibfield  {author} {\bibinfo {author} {\bibfnamefont {T.}~\bibnamefont
  {Renstr\o{}m}}, \bibinfo {author} {\bibfnamefont {H.}~\bibnamefont
  {Utsunomiya}}, \bibinfo {author} {\bibfnamefont {H.~T.}\ \bibnamefont
  {Nyhus}}, \bibinfo {author} {\bibfnamefont {A.~C.}\ \bibnamefont {Larsen}},
  \bibinfo {author} {\bibfnamefont {M.}~\bibnamefont {Guttormsen}}, \bibinfo
  {author} {\bibfnamefont {G.~M.}\ \bibnamefont {Tveten}}, \bibinfo {author}
  {\bibfnamefont {D.~M.}\ \bibnamefont {Filipescu}}, \bibinfo {author}
  {\bibfnamefont {I.}~\bibnamefont {Gheorghe}}, \bibinfo {author}
  {\bibfnamefont {S.}~\bibnamefont {Goriely}}, \bibinfo {author} {\bibfnamefont
  {S.}~\bibnamefont {Hilaire}}, \bibinfo {author} {\bibfnamefont {Y.-W.}\
  \bibnamefont {Lui}}, \bibinfo {author} {\bibfnamefont {J.~E.}\ \bibnamefont
  {Midtb\o{}}}, \bibinfo {author} {\bibfnamefont {S.}~\bibnamefont {Peru}},
  \bibinfo {author} {\bibfnamefont {T.}~\bibnamefont {Shima}}, \bibinfo
  {author} {\bibfnamefont {S.}~\bibnamefont {Siem}}, \ and\ \bibinfo {author}
  {\bibfnamefont {O.}~\bibnamefont {Tesileanu}},\ }\href@noop {} {\bibinfo
  {journal} {arXiv:1804.07654 [nucl-ex]}\ }\BibitemShut {NoStop}%
\bibitem [{\citenamefont {Filipescu}\ \emph {et~al.}(2014)\citenamefont
  {Filipescu}, \citenamefont {Gheorghe}, \citenamefont {Utsunomiya},
  \citenamefont {Goriely}, \citenamefont {Renstr\o{}m}, \citenamefont {Nyhus},
  \citenamefont {Tesileanu}, \citenamefont {Glodariu}, \citenamefont {Shima},
  \citenamefont {Takahisa}, \citenamefont {Miyamoto}, \citenamefont {Lui},
  \citenamefont {Hilaire}, \citenamefont {P\'eru}, \citenamefont {Martini},\
  and\ \citenamefont {Koning}}]{filipescu}%
  \BibitemOpen
\bibfield  {journal} {  }\bibfield  {author} {\bibinfo {author} {\bibfnamefont
  {D.~M.}\ \bibnamefont {Filipescu}}, \bibinfo {author} {\bibfnamefont
  {I.}~\bibnamefont {Gheorghe}}, \bibinfo {author} {\bibfnamefont
  {H.}~\bibnamefont {Utsunomiya}}, \bibinfo {author} {\bibfnamefont
  {S.}~\bibnamefont {Goriely}}, \bibinfo {author} {\bibfnamefont
  {T.}~\bibnamefont {Renstr\o{}m}}, \bibinfo {author} {\bibfnamefont {H.-T.}\
  \bibnamefont {Nyhus}}, \bibinfo {author} {\bibfnamefont {O.}~\bibnamefont
  {Tesileanu}}, \bibinfo {author} {\bibfnamefont {T.}~\bibnamefont {Glodariu}},
  \bibinfo {author} {\bibfnamefont {T.}~\bibnamefont {Shima}}, \bibinfo
  {author} {\bibfnamefont {K.}~\bibnamefont {Takahisa}}, \bibinfo {author}
  {\bibfnamefont {S.}~\bibnamefont {Miyamoto}}, \bibinfo {author}
  {\bibfnamefont {Y.-W.}\ \bibnamefont {Lui}}, \bibinfo {author} {\bibfnamefont
  {S.}~\bibnamefont {Hilaire}}, \bibinfo {author} {\bibfnamefont
  {S.}~\bibnamefont {P\'eru}}, \bibinfo {author} {\bibfnamefont
  {M.}~\bibnamefont {Martini}}, \ and\ \bibinfo {author} {\bibfnamefont
  {A.~J.}\ \bibnamefont {Koning}},\ }\href {\doibase
  10.1103/PhysRevC.90.064616} {\bibfield  {journal} {\bibinfo  {journal} {Phys.
  Rev. C}\ }\textbf {\bibinfo {volume} {90}},\ \bibinfo {pages} {064616}
  (\bibinfo {year} {2014})}\BibitemShut {NoStop}%
\bibitem [{\citenamefont {Capote~{\it{et al.}}}(2019)}]{ripl3}%
  \BibitemOpen
  \bibfield  {author} {\bibinfo {author} {\bibfnamefont {R.}~\bibnamefont
  {Capote~{\it{et al.}}}},\ }\href@noop {} {\bibfield  {journal} {\bibinfo
  {journal} {Reference Input Parameter Library, RIPL-2 and RIPL-3, available
  online at http://www-nds.iaea.org/RIPL-3/}\ } (\bibinfo {year}
  {2019})}\BibitemShut {NoStop}%
\bibitem [{\citenamefont {Brown}\ and\ \citenamefont {Rae}(2014)}]{bro14Nu}%
  \BibitemOpen
  \bibfield  {author} {\bibinfo {author} {\bibfnamefont {B.~A.}\ \bibnamefont
  {Brown}}\ and\ \bibinfo {author} {\bibfnamefont {W.~D.~M.}\ \bibnamefont
  {Rae}},\ }\href@noop {} {\bibfield  {journal} {\bibinfo  {journal} {Nucl.
  Data Sheets}\ }\textbf {\bibinfo {volume} {120}},\ \bibinfo {pages} {115}
  (\bibinfo {year} {2014})}\BibitemShut {NoStop}%
\bibitem [{\citenamefont {Cooper}\ \emph {et~al.}(2018)\citenamefont {Cooper},
  \citenamefont {Beausang}, \citenamefont {Humby}, \citenamefont {Simon},
  \citenamefont {Burke}, \citenamefont {Hughes}, \citenamefont {Ota},
  \citenamefont {Reingold}, \citenamefont {Saastamoinen},\ and\ \citenamefont
  {Wilson}}]{cooper18}%
  \BibitemOpen
  \bibfield  {author} {\bibinfo {author} {\bibfnamefont {N.}~\bibnamefont
  {Cooper}}, \bibinfo {author} {\bibfnamefont {C.~W.}\ \bibnamefont
  {Beausang}}, \bibinfo {author} {\bibfnamefont {P.}~\bibnamefont {Humby}},
  \bibinfo {author} {\bibfnamefont {A.}~\bibnamefont {Simon}}, \bibinfo
  {author} {\bibfnamefont {J.~T.}\ \bibnamefont {Burke}}, \bibinfo {author}
  {\bibfnamefont {R.~O.}\ \bibnamefont {Hughes}}, \bibinfo {author}
  {\bibfnamefont {S.}~\bibnamefont {Ota}}, \bibinfo {author} {\bibfnamefont
  {C.}~\bibnamefont {Reingold}}, \bibinfo {author} {\bibfnamefont
  {A.}~\bibnamefont {Saastamoinen}}, \ and\ \bibinfo {author} {\bibfnamefont
  {E.}~\bibnamefont {Wilson}},\ }\href {\doibase 10.1103/PhysRevC.98.044618}
  {\bibfield  {journal} {\bibinfo  {journal} {Phys. Rev. C}\ }\textbf {\bibinfo
  {volume} {98}},\ \bibinfo {pages} {044618} (\bibinfo {year}
  {2018})}\BibitemShut {NoStop}%
\bibitem [{\citenamefont {Sieja}(2018)}]{sieja2018}%
  \BibitemOpen
  \bibfield  {author} {\bibinfo {author} {\bibfnamefont {K.}~\bibnamefont
  {Sieja}},\ }\href {\doibase 10.1103/PhysRevC.98.064312} {\bibfield  {journal}
  {\bibinfo  {journal} {Phys. Rev. C}\ }\textbf {\bibinfo {volume} {98}},\
  \bibinfo {pages} {064312} (\bibinfo {year} {2018})}\BibitemShut {NoStop}%
\bibitem [{\citenamefont {Midtb\o{}}\ \emph {et~al.}(2018)\citenamefont
  {Midtb\o{}}, \citenamefont {Larsen}, \citenamefont {Renstr\o{}m},
  \citenamefont {Bello~Garrote},\ and\ \citenamefont {Lima}}]{mid18}%
  \BibitemOpen
  \bibfield  {author} {\bibinfo {author} {\bibfnamefont {J.~E.}\ \bibnamefont
  {Midtb\o{}}}, \bibinfo {author} {\bibfnamefont {A.~C.}\ \bibnamefont
  {Larsen}}, \bibinfo {author} {\bibfnamefont {T.}~\bibnamefont {Renstr\o{}m}},
  \bibinfo {author} {\bibfnamefont {F.~L.}\ \bibnamefont {Bello~Garrote}}, \
  and\ \bibinfo {author} {\bibfnamefont {E.}~\bibnamefont {Lima}},\ }\href
  {\doibase 10.1103/PhysRevC.98.064321} {\bibfield  {journal} {\bibinfo
  {journal} {Phys. Rev. C}\ }\textbf {\bibinfo {volume} {98}},\ \bibinfo
  {pages} {064321} (\bibinfo {year} {2018})}\BibitemShut {NoStop}%
\end{thebibliography}
\end{document}